\documentclass[a4paper,11pt]{article}
\usepackage{jheppub} 
\usepackage{lineno}
\usepackage{lineno}
\usepackage{subfig}
\usepackage{verbatim}
\usepackage{amsmath}
\usepackage{graphicx}
\usepackage{physics}
\usepackage{orcidlink}

\title{\boldmath A Note on Chaos in Hayward Black Holes with String Fluids}

\author[a]{Aditya Singh\,\orcidlink{0000-0002-2719-5608}}
\affiliation[a]{Department of Physics, Indian Institute of Technology (Indian School of Mines) Dhanbad, \\Jharkhand 826004, India.}
\author[a]{, Ashes Modak\,\orcidlink{0009-0005-4392-7337}}
\author[a]{, Binata Panda\,\orcidlink{0000-0001-9960-8224}}

\emailAdd{24pr0148@iitism.ac.in}
\emailAdd{22dr0061@iitism.ac.in}
\emailAdd{binata@iitism.ac.in}

\abstract{
    In this work, we first examine the onset of thermodynamic chaos in Hayward AdS black holes with string fluids, emphasizing the effects of temporal and spatially periodic perturbations. We apply Melnikov's approach to examine the perturbed Hamiltonian dynamics and detect the onset of chaotic behavior. In the case of temporal perturbations induced by thermal quenches, chaos occurs for perturbation amplitude $\gamma$ exceeding a critical threshold, determined by charge $q$ and the string fluid parameter. From the equation of state of the black hole, a general condition is established indicating that under temporal perturbations, the existence of charge is an essential prerequisite for chaos. However, regardless of the presence of charge, spatial perturbations result in chaotic behavior.
    Further next, we compute the Lyapunov exponent associated with the thermodynamic system to further quantify chaotic behavior beyond the threshold condition. We demonstrate that the string fluid density and the Hayward regularization parameter have a considerable effect on the amplitude of the Lyapunov exponent, showing the control of thermal instability by regular geometry corrections and matter sources. 
    These results highlight the rich nonlinear dynamics arising from the interplay of geometric regularization, matter content, and phase-space instability.}


\begin{document}
\maketitle
\flushbottom

\section{Introduction}

Black hole solutions in general relativity exhibit rich thermodynamic behavior, revealing deep connections between gravitation, quantum theory and statistical mechanics. Of particular interest are black hole phase transitions in various spacetime backgrounds, notably in asymptotically anti–de Sitter (AdS) spacetimes, which have been extensively studied both from a gravitational perspective and within the framework of holographic dualities~\cite{Bekenstein:1973ur,Bekenstein:1974ax,Hawking:1976de,Hawking:1982dh,Maldacena:1997re,Witten:1998qj}. Among these, regular black hole solutions such as the Hayward black hole, which incorporate nonlinear matter sources like string fluids, provide a compelling framework to explore singularity-free geometries and their rich thermodynamic behavior. Recently, extended phase space thermodynamics has been developed by reinterpreting the cosmological constant as a dynamical thermodynamic variable related to pressure~\cite{Kastor:2009wy,Sekiwa:2006qj,LarranagaRubio:2007ut,Kubiznak2012}. A pressure-volume ($PdV$) work term has been introduced to the first law of black hole mechanics within this paradigm, enabling a more thorough thermodynamic description~\cite{Dolan:2010ha,Cvetic:2010jb,Dolan:2011xt,Teitelboim:1985dp,HENNEAUX1984415,Henneaux:1989zc}. Black hole phase transitions, i.e. transitions between small and large black hole phases, are precisely related to the classical liquid–gas transitions described by the van der Waals equation of state, in line with studies of the pressure–volume ($P-V$) criticality across various black hole spacetimes~\cite{Chamblin:1999tk,Kubiznak2012,Kubiznak2017,Gunasekaran2012,Singh:2020tkf,Singh:2023hit,Singh:2023ufh,Singh:2025ueu,Anand:2025mlc,Anand:2025vfj}.

\par It is well known that chaotic dynamics can arise in a wide class of nonlinear systems, including cosmological models and gravitational settings such as black hole spacetimes. From the study of quasinormal mode spectra in charged and higher curvature black holes, such as Reissner-Nordström and Gauss–Bonnet solutions, to the analysis of Lyapunov exponents determining orbital stability, black hole physics has used a variety of methods to investigate chaos~\cite{Bombelli:1991eg,Letelier:1996he,Santoprete:2001wz,Monerat:1998tw,Asano:2016qsv,Dubeibe:2007hba,Gair:2007kr,Chen:2016tmr,Cardoso:2008bp,GuckenheimerHolmes1983,Chabab:2018lzf}. Furthermore, the Melnikov approach, a standard method that identifies homoclinic bifurcations and the onset of chaos in perturbed Hamiltonian systems, has been applied to dynamical perturbations and geodesic motion in black hole backgrounds~\cite{10.1098/rsta.1979.0068,Holmes1981,HOLMES1990137,SLEMROD1985135}. The fundamental and prevalent characteristic of nonlinear dynamical systems is chaos, often characterized by a significant sensitivity to initial conditions. Despite the seemingly random evolution of such systems, they often display underlying features including self-similarity, inter-connectivity, and observable patterns. A familiar aspects of chaos is the butterfly effect, which posits that even infinitesimally small changes in the initial state of the system can lead to exponentially divergent outcomes, impacting the final configuration of the system and observable quantities. Additionally, this unforeseen behavior may occur even in deterministic systems, which are devoid of inherent randomness and whose future evolution is exclusively dictated by their initial conditions. The resulting phenomenon, referred to as deterministic chaos or classical chaos, has been thoroughly investigated using a variety of mathematical techniques, including as Melnikov functions, Lyapunov exponents, and Kolmogorov-Sinai (KS) entropy~\cite{FELDERHOF1970541,Widom1977,smale1963diffeomorphisms,Pesin:1997yb}.

\par It is interesting to note that chaos-like characteristics are also present in quantum systems regulated by nonlinear dynamics, however they appear differently because of the essentially a distinct structure of quantum mechanics. Chaos is usually investigated in the quantum regime through approaches ranging from random matrix theory to semiclassical studies (see Refs.\cite{Chabab:2018lzf,Pesin:1997yb,10.1098/rsta.1979.0068,Holmes1981,HOLMES1990137,SLEMROD1985135,FELDERHOF1970541,Widom1977,smale1963diffeomorphisms,DeFalco:2020yys,Larsen:1993nt,Frolov:1999pj}. Even with these substantial tools, a comprehensive and widely accepted overview of quantum chaos remains elusive. A deeper insight into the nature of quantum chaos has emerged from the AdS/CFT correspondence. It has been demonstrated that out-of-time-order correlators (OTOCs) are useful probes of quantum chaotic behavior in the dual boundary field theory. A remarkable bulk-boundary duality has been established in the bulk description, where the collision of gravitational shock-waves close to the event horizon of asymptotically AdS black holes has been demonstrated to resemble the behavior of OTOCs~\cite{SLEMROD1985135,FELDERHOF1970541,Widom1977,smale1963diffeomorphisms,Pesin:1997yb,DeFalco:2020yys,Larsen:1993nt,Frolov:1999pj,Bronnikov:1994ja,Karan:2023hfk,Chakrabortty:2022kvq,PhysRevLett.55.2656,Sen:2005iz,Qaemmaqami:2017bdn}. These shockwave analyses yield two important indicators of chaos, the butterfly velocity $v_{B}$, which measures the spatial spread of these perturbations through spacetime, and the Lyapunov exponent $\lambda$, which quantifies the sensitivity to initial perturbations in quantum systems (analogous to the classical Lyapunov exponent)~\cite{Shukla:2024tkw,Yang:2025fvm,Gogoi:2024akv}. In addition to defining the level of chaos, these serve as a link between classical and quantum insights of dynamical instability.


\par The term ``classical chaos" describes the complex emergent behavior of deterministic systems whose time evolution is nearly unpredictable beyond a finite timescale, yet being regulated by well-defined equations of motion. This unpredictability, which significantly reduces the accuracy of long-term forecasts, results from an exponential sensitivity to initial conditions rather than from inherent randomness. Three basic factors govern the time horizon across which forecasts continue to be valid: the level of uncertainty permissible in the forecast, the accuracy of the initial condition measurement, and a characteristic timescale called the Lyapunov time. Chaos can also occasionally result from the intrinsic complexity of the Hamiltonian of the system, particularly in systems with multiple degrees of freedom. In such systems, individual highly excited energy eigenstates are thought to mimic thermal ensembles, according to the Eigenstate Thermalization Hypothesis (ETH)~\cite{Deutsch:1991msp,Srednicki:1994mfb}. Without explicit external stochasticity, these systems can therefore display chaotic dynamics and thermalization. This thermodynamic analogy becomes particularly profound in the context of black hole physics, especially in the semiclassical regime.

\par The Hayward black holes with string fluids, which are normal black hole solutions that are generated by coupling gravity to nonlinear string-inspired matter fields, are particularly well-suited for investigating both classical and thermal chaos. These black holes evade singularities through a modified core geometry and support rich thermodynamic and dynamical structures due to the presence of effective matter content resembling a string cloud. Hayward black holes with string fluids show signs of thermal chaos in various regimes of the parameter space, consistent with holographic considerations and the ETH assumptions. Both at the classical and semiclassical levels, their nonlinear field content and modified spacetime structure results in nontrivial dynamics, which enable the analysis of thermal instability, phase space mixing, and Lyapunov exponents in a gravitational setting. The investigation of chaotic dynamics in such regular black hole spacetimes enhances our knowledge of classical chaos in gravitational systems and discusses the relation of geometry, thermodynamics, and microscopic degrees of freedom in quantum gravity, as will be covered in the following sections.

A complementary and increasingly useful way to probe black hole phase transitions is through Lyapunov exponents associated with unstable circular geodesics. In black hole spacetimes, Lyapunov exponents quantify the local instability of null and timelike trajectories and therefore provide a direct dynamical measure of orbital sensitivity to perturbations. Recent studies~\cite{Cardoso:2008bp,Yang:2023hci,Yang:2025fvm} have shown that this quantity is not merely a kinematical indicator of geodesic instability, but also encodes important thermodynamic information. In the present work, we incorporate this Lyapunov analysis as a diagnostic complementary to the Melnikov method. The roles of the two approaches are conceptually distinct. Melnikov's method detects the onset of chaos in the perturbed thermodynamic Hamiltonian system through the transverse splitting of homoclinic or heteroclinic structures, thereby providing a global criterion for chaotic behavior in the extended phase space. By contrast, the Lyapunov exponent probes the local instability of circular null and timelike geodesics and reveals how this instability is organized across the thermodynamic branches and near the phase transition~\cite{Guo:2022kio,Bezboruah:2025udi}. Studying these two approach together therefore allows us to connect the global chaotic structure of the perturbed thermodynamic phase space with the local dynamical instability of the underlying black hole geometry. This combined perspective is particularly useful for Hayward-AdS black holes with string fluids, where nonlinear matter effects modify both the thermodynamic phase structure and the geodesic instability pattern.

\quad {\bf Motivations:} In this work, we extend the investigation of chaotic dynamics in black hole thermodynamics by focusing on Hayward black holes coupled to string fluids within the framework of extended phase space. These regular black hole solutions are motivated by effective field theories derived from quantum gravity and string theory and originate from modifications to the matter sector in Einstein gravity. The inclusion of a string fluid alters both the geometric and the thermodynamic structure of the black hole, introducing additional parameters that enrich the phase space and allow for novel dynamical behavior. In black holes exhibiting van der Waals like phase transitions, such configurations are particularly interesting because they offer an ideal environment for investigating the onset of chaos near criticality. Although instabilities of geodesics and chaotic dynamics of test particles have been investigated in a variety of regular and singular black hole backgrounds, the thermodynamic origin of chaos in regular black holes has not received as much attention. The extended thermodynamic framework enables a more complete analogy with fluid systems and allows the application of tools such as Melnikov’s approach to detect chaotic behavior under perturbations in the $P-V$ sector. In addition, the Lyapunov exponent offers a complementary dynamical probe by quantifying the local instability of circular null and timelike geodesics. This makes it possible to examine whether the phase structure and thermodynamic branches of the black hole are also reflected in the corresponding geodesic instability. Based on previous studies of chaos in charged AdS black holes, our goal is to ascertain if chaos in thermodynamic phase space is a general characteristic of systems exhibiting van der Waals type criticality, even when singularities are not present~\cite{Cai:2013qga,Lan:2015bia,Kim:2013xu,Belhaj:2012bg}. Due to its regular core structure and the presence of an extra coupling parameter that characterizes the string fluid, the Hayward black hole with a string fluid constitutes an ideal test case. For periodic perturbations, our study indicates that the system enters a chaotic domain beyond a specific bound involving this coupling parameter.

\quad The paper is organized as follows: In section-(\ref{Melnikov's Method}), we describe Melnikov's approach to chaotic dynamics. Further, we review the basic thermodynamics and properties of Hayward black holes with string fluids in extended phase space. Section-(\ref{Temporal chaos}) describes the influence of a small periodic perturbation on temporal chaos for the system. In section-(\ref{Spatial chaos}), we discuss the emergence of intricate spatial structures arising from nonlinearities and dynamical instabilities. Section-(\ref{Lyapunov}) is focused on analyzing phase transition and thermodynamic stability of Hayward black holes using Lyapunov exponent. Finally, we end our discussion in section-(\ref{Remarks}) with remarks on the chaotic dynamics of Hayward AdS black holes with string fluids.  


\section{Melnikov's Method and its Application to Chaotic Dynamics}\label{Melnikov's Method}

We examine the Melnikov method in this subsection, which is a fundamental analytical method for examining the onset of chaotic dynamics in nearly integrable Hamiltonian systems. This method analyzes the effect of small time-periodic perturbations on saddle-type fixed point invariant manifolds. Consider the dynamical system described by the equation
\begin{equation}\label{evolution}
\dot{Z} = f(Z) + \varepsilon g(Z,t), \quad Z \in \mathbb{R}^{2n}, \quad \varepsilon \ll 1 \ ,
\end{equation}
where $f(Z)$ generates a Hamiltonian flow and $g(Z,t)$ introduces a smooth, time-periodic perturbation. A homoclinic trajectory connecting a hyperbolic saddle point to itself is admitted by the unperturbed system ($\varepsilon = 0$). The splitting of the corresponding stable and unstable manifolds as a result of the perturbation can be measured by the Melnikov method. Although homoclinic connections are frequently used to demonstrate the theory, it may also be applied to heteroclinic orbits between different saddle points, even if the explicit orbit $Z_0(t)$ is unknown.

When perturbation takes place, the stable and unstable manifolds overlap transversely at an infinite number of locations because they no longer coincide.   This indicates sensitivity to initial conditions, thereby rendering it an intrinsic sign of chaos. The transverse separation between these manifolds at time $t_0$ is proportional to the Melnikov function $M(t_0)$, with the distance given by
\begin{equation}
    d(t_0) = \frac{\varepsilon M(t_0)}{|\mathbf{f}(Z_0(0))|} \ .
\end{equation}
The Melnikov function can be computed as
\begin{equation}\label{melnikov}
M(t_0) = \int_{-\infty}^{\infty} f^T\left(Z_0(t - t_0)\right) \Omega_n g\left(Z_0(t - t_0), t\right) dt \ ,
\end{equation}
where $\Omega_n$ denotes the canonical symplectic matrix\footnote{In Hamiltonian mechanics, the dynamics of a system with $n$ degrees of freedom is described in phase space using
\begin{equation*}
    Z = (Q^1, \dots, Q^n, \mathcal{P}_1, \dots, \mathcal{P}_n)^\top \in \mathbb{R}^{2n}
\end{equation*}
The evolution is governed by a Hamiltonian function $\mathcal{H}(x, t)$, representing the total energy. The equations of motion can be written as 
\begin{equation*}
\dot{Q}^i = \frac{\partial \mathcal{H}}{\partial \mathcal{P}_i}, \quad \dot{\mathcal{P}}_i = -\,\frac{\partial \mathcal{H}}{\partial Q^i}.
\end{equation*}
compactly written using the canonical symplectic matrix $\Omega$ as 
\begin{equation*}
\dot{Z} = 
\begin{pmatrix}
0 & \mathcal{I} \\
- \mathcal{I} & 0
\end{pmatrix}
\begin{pmatrix}
\partial \mathcal{H} / \partial \dot{Q_i} \\
\partial \mathcal{H} / \partial \dot{\mathcal{P}_i}
\end{pmatrix}
=
\begin{pmatrix}
\partial \mathcal{H} / \partial \dot{\mathcal{P}_i} \\
- \partial \mathcal{H} / \partial \dot{Q_i} 
\end{pmatrix} \;\;\;\;\;\text{finally}\;\;\; \boxed{\dot{Z} = \Omega \, \nabla \mathcal{H}}
\end{equation*}} in a Hamiltonian system with $m$ degrees of freedom, it has the general block form
$$
\Omega_n =
\begin{pmatrix}
0 & \mathcal{I}_n \\
- \mathcal{I}_n & 0
\end{pmatrix},
$$

where $\mathcal{I}_n$ is the $n \times n$ identity matrix and the zero entries denote $n \times n$ zero matrices. The number of degrees of freedom in the perturbation is represented by the dimensional index $n$. The distance between the manifolds is quantified by the function $M(t_0)$; a simple zero for this function means that the manifolds intersect transversely at time $t_0$. According to the Smale–Birkhoff theorem \cite{smale1963diffeomorphisms,Bronnikov:1994ja}, such crossings indicate chaotic behavior, which ensures the existence of symbolic dynamics identical to a Smale horseshoe.


\subsection{Review of Hayward Black Holes Surrounded by String Fluids}\label{Hayward Review}
In this section, we give a quick overview of Hayward black holes with string fluids. We begin with investigating the Hayward black hole spacetime \cite{Hayward:2005gi,Nascimento:2023tgw}, which emerges as a regular solution to Einstein's field equations when coupled to a nonlinear electromagnetic field. A nonlinear electrodynamics (NED) arrangement controls the gravitational source in this configuration \cite{Belhaj:2012bg,Hayward:2005gi,Nascimento:2023tgw,Molina:2021hgx,Bronnikov:2000vy}. The action describing the system, wherein the nonlinear electromagnetic field is minimally coupled to gravity, can be expressed as~\cite{Molina:2021hgx,Bronnikov:2000vy},
\begin{equation}
\mathcal{S} = \frac{1}{16\pi}\int d^4x\sqrt{-g} \left[R-2\Lambda+\mathcal{L}(\mathcal{F}) \right]+S_{M}, \;\;\;  \;\;\mathcal{L}(\mathcal{F}) = \frac{12 \sqrt{2}\mathrm{h}\mathcal{F}^{3/2}}{ \kappa^2 \left(1+\left(2\mathrm{h}^2\mathcal{F}\right)^{3/4}\right)^2}
\label{eq_action}
\end{equation}
where $R$ represents the Ricci scalar, $g$ is the determinant of the spacetime metric $g_{\mu\nu}$,  $\mathcal{F}=F^{\mu\nu}F_{\mu\nu}$ is the electromagnetic field invariant, $\mathrm{h}$ is the Hayward parameter, usually connected to a fundamental length scale, and $\kappa^2 = 8\pi$. $S_{M}$ is the action of the fluid of strings. The modified Einstein field equations are derived by performing a variation of the action \eqref{eq_action} with respect to the metric tensor\cite{Molina:2021hgx,Bronnikov:2000vy,Fan:2016hvf},
\begin{equation}
R_{\mu\nu} - \frac{1}{2}\,g_{\mu\nu}R-\Lambda g_{\mu\nu} = 2\frac{\partial\mathcal{L}(\mathcal{F})}{\partial F}F_{\mu\sigma}F^{\sigma}{_\nu}-\frac{1}{2}g_{\mu\nu}\mathcal{L}(\mathcal{F})+T_{\mu\nu}^{FS} .
\label{Einstein eqn}
\end{equation}
In a spherically symmetric spacetime with a purely magnetic configuration, according to~\cite{Bronnikov:2000vy}, the electromagnetic field tensor \( F_{\mu\nu} \) possesses a single non-vanishing component, which can be expressed as follows, $F_{23} = q_m \sin{\theta}$, leading to the field invariant,
\begin{equation}\label{F}
    F = \frac{2q_m^2}{r^4}.
\end{equation}
Hayward proposed that the parameter $h$, characterizing the regularization scale, is of the order of the Planck length and can be related to the magnetic charge $q_m$ via the relation \cite{Fan:2016hvf,Toshmatov:2018cks,Bronnikov:2017tnz}
\begin{equation}
q_m = \frac{\sqrt[3]{4m^2 h}}{2} \ ,
\label{magnetic charge}
\end{equation}
With this setup, the components of the stress-energy tensor related to the Hayward solution have the structure~\cite{Hayward:2005gi},
\begin{equation}
T_t^{t} = T_r^{r} = \frac{12 m^2 \mathrm{h}^2}{(2m \mathrm{h}^2 + r^3)^2}\;\;\;;\;\;\; T_\theta^{\theta} = T_\phi^{\phi} = -\frac{24m^2\mathrm{h}^2 (r^3 - m\mathrm{h}^2)}{(2m \mathrm{h}^2 + r^3)^3} \ ,
\label{eq:1.6}
\end{equation}
where $h$ and $m$ are positive constants. The EOM \eqref{Einstein eqn} is modified to incorporate two more physical sources: a fluid of strings and a cosmological constant. The string fluid is represented by the term $T_{\mu\nu}^{\text{FS}}$ on the right-hand side and the term $-\Lambda g_{\mu\nu}$ on the left-hand side. For the inclusion of strings fluid, the energy–momentum tensor can be expressed as~\cite{Letelier:1979ej,Soleng:1993yr},
\begin{equation}
(T_{\mu\nu})_{\text{string}} = \frac{1}{\sqrt{-\gamma}} \rho (\Sigma_{\mu\beta} \Sigma^{\beta}_{\;\nu}) \, ,
\label{energy momentum 2}
\end{equation}
where $\rho$ denotes the proper string density, $\rho/\sqrt{-\gamma}$ is the gauge-invariant density, and $\Sigma_{\mu\nu}$ is a surface-forming antisymmetric bivector that characterizes the string configuration. Imposing the conservation of the total energy-momentum tensor $\nabla_{\mu}T^{\mu\nu}=0$, and employing the identity $\Sigma_{\mu\beta} \Sigma^{\beta}_{\;\tau} \Sigma^{\tau}_{\;\nu} = \gamma \Sigma_{\nu\mu} \,$, we get
\begin{equation}
\nabla_\mu(\rho \Sigma^{\mu\beta}) \Sigma^{\nu}_{\;\beta} = 0 \, ,
\end{equation}
In a coordinate system adapted to the surface parameterization of the string cloud, the aforementioned condition reduces to
\begin{equation}\label{sigma2}
\partial_\mu(-g \rho \Sigma^{\mu\beta}) = 0 \, ,
\end{equation}
where both $\rho$ and $\Sigma^{\mu\nu}$ are assumed to depend only on the radial coordinate $r$ consistent with static, spherically symmetric spacetime configurations. In this setting, the non-zero component of the bivector is $\Sigma^{tr}=- \Sigma^{rt}$. As a result, the energy–momentum tensor components reduce to $T^t_{t} = T^r_{r} = -\rho |\Sigma^{tr}|$ and eq.~(\ref{sigma2}) implies $\partial_r(r^2 T^t_{t})=0$. Solving this, the energy–momentum tensor for the cloud of strings is obtained as \cite{Ghosh:2014pga,Ghosh:2014dqa},
\begin{equation}
(T^\nu_\mu)_{\text{string}} = \text{diag} \left( \frac{a}{r^2}, \frac{a}{r^2}, 0, 0 \right) \, ,
\label{energy momentum 3}
\end{equation}
We now write down the general form of the metric for the Hayward-AdS black hole in the presence of a string fluid
\begin{equation}\label{General Metric}
    ds^2=-f(r)dt^2+\frac{dr^2}{f(r)}+r^2 d\Omega^2,
\end{equation}
Solving the Einstein field equations using eq.~(\ref{eq_action}), (\ref{Einstein eqn}), (\ref{F}) and (\ref{energy momentum 3}), the resulting metric function takes the form,
\begin{equation}
f(r) = 1 - a - \frac{2 M r^2}{q^3 + r^3} + \frac{ r^2}{l^{2}} \, ,
\label{fr}
\end{equation}
where $M$ can be interpreted as the ADM mass of the black hole, $a$ is the string fluid parameter, $q$ is the Hayward regularization parameter and $l$ is equal to $(-\frac{3}{\Lambda})^{\frac{1}{2}}$, where $\Lambda$ can be considered as the dynamical cosmological constant related to the thermodynamic pressure $P$ of the black hole as $P=- \Lambda/8\pi$~\cite{Karch:2015rpa,Mancilla:2024spp,Mann:2025xrb,Sood:2022fio,Dymnikova:2001fb}.

\subsection{Thermodynamics of Hayward black holes with string fluids}
\begin{figure}[ht]
	\centering
	\includegraphics[width=7.5cm,height=5cm]{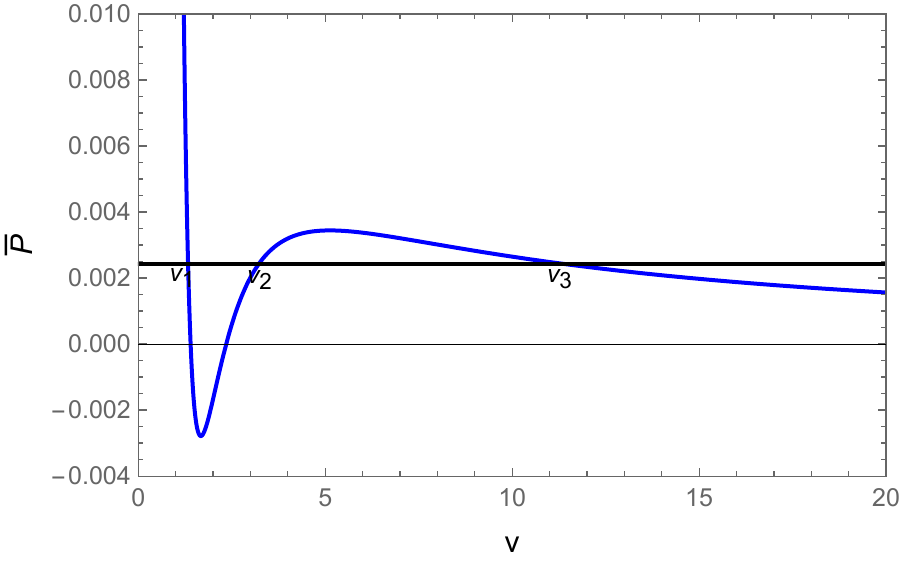}
	\caption{The behavior of $\bar{P}-v$ isotherm for fixed values of temperature $T_{0}=0.036$, string fluid parameter  $a=0.4$ and charge $q=0.4$.}
	\label{case 0}
\end{figure}

Thermodynamics of black holes is a very interesting and rich field of gravitational physics. In the following section, we investigate black hole thermodynamic properties. John Wheeler \cite{Bekenstein:1972tm,Wald:1999vt,Ma:2014qma} first realized that entropy and temperature can be attributed to black holes themselves in order for the generalized second law of thermodynamics to hold in systems with black holes. This insight suggests that an infalling object not only transfers its mass, angular momentum, and electric charge, but also its entropy to the black hole. Later, Bekenstein and Hawking's seminal work transformed the field by establishing evident thermodynamic analogies: they associated temperature with the event horizon's surface gravity and black hole entropy to the event horizon's area~\cite{Bekenstein:1972tm,Wald:1999vt}. The Hawking temperature is associated with surface gravity $\kappa$ by the equation $T=\kappa/2\pi$, where surface gravity $\kappa$ is defined as $\kappa= \frac{f'(r)}{2}|_{r=r_+}$, with $f(r)$ representing the metric function and $r_+$ indicating the radius of the event horizon~\cite{Wald:1999vt}. Applying the above relation and using eq.~(\ref{fr}), we can compute the temperature of the Hayward black holes with string fluids as,
\begin{equation}\label{temperature}
   T = \frac{2 (a-1) q^3 l^2-(a-1) l^2 r_+^3+3 r_+^5}{4 \pi  l^2 r \left(q^3+r_+^3\right)}
\end{equation}
The framework of extended phase space thermodynamics offers a more unified and consistent approach to black hole thermodynamics by identifying the black hole mass $M$ with the spacetime enthalpy and treating the cosmological constant $\Lambda$ as a thermodynamic pressure $P$~\cite{Naseer:2025,Naseer:2025ghn}. Since our black hole solution has two horizons, these horizons must have varying temperatures. Thus, it is theoretically not possible to achieve thermal equilibrium. In our analysis, however, we assume that the inner (Cauchy) horizon stays dynamically decoupled, whereas the event horizon is capable of reaching a quasi-equilibrium configuration when coupled to an external thermal reservoir. The first law of black hole thermodynamics can still be reliably applied at the event horizon to examine thermodynamic behavior under this quasi-equilibrium assumption. In extended phase space, the first law generalizes to the magnetic charge and the cosmological constant, as well as their thermodynamic conjugates, for a static, charged black hole. The modified first can be written as,
\begin{equation}\label{first law}
    dM =TdS + VdP +\phi dq
\end{equation}
where $\phi$ denotes the potential. $S$ is the entropy and $V$ is the volume of the black hole conjugate to the thermodynamic pressure $P$ of the black hole.
Utilizing eq.~\eqref{temperature} and putting $v = 2r_+$, it is straightforward to compute the equation of state,

\begin{equation}\label{equation of state}
\bar{P}(T,v)= \frac{T}{v}-\frac{1}{2 \pi  v^2}+\frac{a}{2 \pi  v^2}+\frac{8 q^3 T}{v^4} -\frac{8 a q^3}{\pi  v^5}+\frac{8 q^3}{\pi  v^5},
\end{equation}
where $v$ denotes specific volume. For temperatures below the critical value $T_c$, the $\bar{P}$--$v$ diagram (Figure~\ref{case 0}) exhibits a characteristic non-monotonic behavior that resembles the Van der Waals fluid.

The isotherm features three distinct regions: a stable small black hole branch at low specific volume, an unstable intermediate branch where $\left(\frac{\partial \bar{P}}{\partial v}\right) > 0$, and a stable large black hole branch at high volume. The intermediate region, known as the spinodal region, corresponds to thermodynamically unstable configurations and is typically associated with phase separation. Dynamically, the presence of saddle points in this region supports the formation of homoclinic and heteroclinic orbits, making it a natural setting for the emergence of chaos under temporal or spatial perturbations. The critical temperature ($T_{cr}$) can be determined following the approach outlined in~\cite{Kumar:2024qon},

\begin{equation}
T_{cr}=\frac{\sqrt[3]{25}(1-a)}{4\pi q\sqrt[3]{(118+48\sqrt{6})}}
\end{equation}

The critical temperature $T_{cr}$ marks the boundary between distinct thermodynamic phases of the black hole system. It corresponds to the point at which the $\bar{P}$--$v$ isotherm develops an inflection point, defined by the conditions $\left( \frac{\partial \bar{P}}{\partial v} \right)_T = 0$ and $\left( \frac{\partial^2 \bar{P}}{\partial v^2} \right)_T = 0$. At this temperature, the distinction between small and large black hole phases vanishes.


\section{Analysis of Temporal Chaos}\label{Temporal chaos}
This section analyzes the impact of a minor, temporally periodic disturbance on the system, specifically when it is quenched into the unstable spinodal region. To analyze this, we first construct the Hamiltonian governing the fluid dynamics, based on the black hole's equation of state. Subsequently, we derive the associated Melnikov function, which encodes critical information regarding the onset of chaotic behavior. As a starting point, we consider a specific volume $v_{0}$ lying on an isotherm characterized by temperature $T_0$, which is then subjected to a periodic fluctuation described as follows,
\begin{equation}\label{Temporal Perturbation}
    T=T_{0}+\varepsilon\gamma \cos(\omega t)\cos \mathcal{M}
\end{equation}
where, $\mathcal{M}$ represents a segment of the black hole modeled as a fluid column with a unit cross-sectional area, bounded between two reference points. $\gamma$ is the perturbation amplitude, $\omega$ is the angular frequency of the fluctuation and $\varepsilon<<1$. The fluid is assumed to flow along the $x$-axis within a tube of unit cross section and constant volume. The system contains a total fluid mass of $(2\pi/s)$ occupying a volume of $(2\pi/s)v_0$, where $s > 0$ is a fixed constant characterizing the density distribution. The details of the setup, along with the underlying assumptions, closely follow those presented in earlier analyses. In the present context, the Hamiltonian that governs the system can be expressed symbolically as follows \cite{SLEMROD1985135}:
\begin{equation}\label{Hamiltonian Fundamental}
    H(x,u;T)=\frac{1}{\pi}\int_{0}^{2\pi}\left[\frac{u^{2}}{2}-\int_{v_{0}}^{v}\bar{P}(v,T)dv+\frac{As^{2}}{2}\left(\frac{\partial v}{\partial \mathcal{M}}\right)^{2}\right]d\mathcal{M}
\end{equation}
To carry out the perturbation analysis, we expand $\bar{P}$ in a Taylor series around the point $(T_0, v_0)$ and retain terms up to the $\mathcal{O}(v^4)$,
\begin{equation}\label{Pressure Perturbation}
\begin{split}
    \bar{P}(T_{0},v_{0})&= \bar{P}(T_{0},v_{0})+ \bar{P}_{v}(T_{0},v_{0})(v-v_{0})+\bar{P}_{T}(T_{0},v_{0})(T-T_{0})
   +\frac{\bar{P}_{vvv}}{3!}(T_{0},v_{0})(v-v_{0})^{3}\\
    & +\bar{P}_{vT}(T_{0},v_{0})(v-v_{0})(T-T_0)+\frac{\bar{P}_{vvT}}{2!}(T_{0},v_{0})(v-v_{0})^{2}(T-T_{0})\\
    \end{split}
\end{equation}
In the neighborhood of the inflection point, the functions $v(\mathcal{M},t)$ and $u(\mathcal{M},t)$ can be expressed as Fourier series over the domain $\mathcal{M} \in [0, 2\pi]$\footnote{Where $\mathcal{M}$ represents the mass of a portion of the black hole fluid that has a cross-sectional area of one unit, extending from the reference particle at
$x=0$ to a fluid element at position
$x$, and is computed as follows,
$$\mathcal{M}(x,t)=\int_{0}^{x}\rho(\tilde{x},t)d\tilde{x}$$
Here, $\rho(x,t)$ denotes the fluid density at position $x$ and time $t$. It follows directly from the above relation that $x$ can be expressed as a function of mass: $x = x(\mathcal{M},t)$. We define the fluid velocity as $u(\mathcal{M},t) = \frac{\partial x}{\partial t}(\mathcal{M},t)$.
},

\begin{equation}\label{Fourier v}
v(\mathcal{M},t)=v_{0}+\sum_{k=1}^{\infty}x_{k}(t)\cos(k\mathcal{M}),\qquad
u(\mathcal{M},t)=\sum_{k=1}^{\infty}u_{k}(t)\cos(k\mathcal{M})
\end{equation}
Here $\bar{P}(v,T)$ is the effective equation of state defined in eq.~\eqref{equation of state}. In eq.~\eqref{Hamiltonian Fundamental}, if we substitute eq.~\eqref{Temporal Perturbation}, \eqref{Pressure Perturbation}, \eqref{Fourier v}  we get
\begin{equation}
    H(x,u)=H_{2}(x,u)+R(x,u),
\end{equation}
where $R(x, u)$ denotes the remainder term, which disappears if $x_n = u_n = 0$ for all $n \geq 3$. By neglecting the higher-order modes ($n \geq 3$), the system reduces to a Hamiltonian system with two degrees of freedom, given as \cite{MahishBhamidipati2019},
\begin{equation}
\begin{split}
    H_{2}(x,u)=&\frac{u_{1}^{2}+u_{2}^{2}}{2}-\frac{\bar{P}_{v}}{2}(T_{0},v_{0})(x_{1}^{2}+x_{2}^{2})-\frac{\bar{P}_{vv}}{4}(T_{0},v_{0})x_{1}^{2}x_{2}-\frac{\bar{P}_{vvv}}{32}(T_{0},v_{0})(x_{1}^{4}+x_{4}^{2}+4x_{1}^{2}x_{2}^{2})\\
    &-\bar{P}_{T}(T_{0},v_{0})\varepsilon\gamma \cos(\omega t)x_{1}-\frac{\bar{P}_{vT}}{2}(T_{0},v_{0})\varepsilon\gamma \cos(\omega t)x_{1}x_{2}\\
    &-\frac{\bar{P}_{vvT}}{8}(T_{0},v_{0})\varepsilon\gamma \cos(\omega t)x_{1}(x_{1}^{2}+2x_{2}^{2})+\frac{As^{2}}{2}(T_{0},v_{0})(x_{1}^{2}+4x_{2}^{2})\\
    \end{split}
\end{equation}
The aforementioned form of the Hamiltonian generally holds for all black holes within the extended phase space. In the specific case of a Hayward black hole surrounded by fluid of strings in AdS spacetime, the corresponding $\bar{P}_{v}$, $\bar{P}_{vv}$, $\bar{P}_{vvv}$, $\bar{P}_{T}$, $\bar{P}_{vT}$, $\bar{P}_{vvT}$\footnote{Here, the subscripts of $\bar{P}$ indicate derivatives with respect to the corresponding variables.} is given by,
\begin{equation}
\begin{split}
    \bar{P}_{v} &= \frac{8 q^3 (5 a-4 \pi  T v-5)-v^3 (a+\pi  T v-1)}{\pi  v^6}, \;\;\;\;\;\;\;\;\;\;\; \bar{P}_{vT}= -\frac{32 q^3+v^3}{v^5}\\
    \bar{P}_{vv} &= \frac{v^3 (3 a+2 \pi  T v-3)-80 q^3 (3 a-2 \pi  T v-3)}{\pi  v^7}, \;\;\;\;\;\;\;  \bar{P}_{vvT}=\frac{2 \left(80 q^3+v^3\right)}{v^6}\\ 
    \bar{P}_{vvv} &= \frac{240 q^3 (7 a-4 \pi  T v-7)-6 v^3 (2 a+\pi  T v-2)}{\pi  v^8}, \quad  \bar{P}_{T}= \frac{8 q^3+v^3}{v^4}
\end{split}
\end{equation}
On substituting this, we can write the final expression for the Hamiltonian as,
\begin{equation}
\begin{split}
H_{2}(x,u) = \frac{1}{16}\Bigg[
&-\frac{8\left(8 q^{3}(5a-4\pi T v-5)-v^{3}(a+\pi T v-1)\right)
\left(x_{1}^{2}+x_{2}^{2}\right)}{\pi v^{6}}  \\[4pt]
&-\frac{4\left(v^{3}(3a+2\pi T v-3)-80 q^{3}(3a-2\pi T v-3)\right)
x_{1}^{2}x_{2}}{\pi v^{7}} \\[4pt]
&-\frac{3\left(40 q^{3}(7a-4\pi T v-7)-6 v^{3}(2a+\pi T v-2)\right)
\left(x_{1}^{4}+4x_{1}^{2}x_{2}^{2}+x_{2}^{4}\right)}{\pi v^{8}} \\[4pt]
&+8 A s^{2}\left(x_{1}^{2}+4x_{2}^{2}\right) \\[4pt]
&-\frac{16\left(8 q^{3}+v^{3}\right)x_{1}\gamma\varepsilon\cos(t\omega)}{v^{4}}
+\frac{8\left(32 q^{3}+v^{3}\right)x_{1}x_{2}\gamma\varepsilon\cos(t\omega)}{v^{5}} \\[4pt]
&-\frac{4\left(80 q^{3}+v^{3}\right)x_{1}\left(x_{1}^{2}+2x_{2}^{2}\right)
\gamma\varepsilon\cos(t\omega)}{v^{6}}
+8\left(u_{1}^{2}+u_{2}^{2}\right)
\Bigg].
\end{split}
\end{equation}

where, $(x_{1}, x_{2})$ and $(u_{1}, u_{2})$ denote the position and velocity components of the first two dynamical modes, respectively. The associated equations of motion can be given as,
\begin{equation}\label{x dot}
   \dot{x}_{1}=\frac{\partial H_{2}}{\partial u_{1}}=u_{1}, \quad \dot{x}_{2}=\frac{\partial H_{2}}{\partial u_{2}}=u_{2}
\end{equation}
\begin{equation}\label{u dots}
    \dot{u}_{1}=-\frac{\partial H_{2}}{\partial x_{1}}-\varepsilon\mu_{0}su_{1}, \quad  \dot{u}_{2}=-\frac{\partial H_{2}}{\partial x_{2}}-4\varepsilon\mu_{0}su_{2}
\end{equation}
In the case we are considering, the value of $\dot{u}_{1}$ and $\dot{u}_{2}$ can be given as \footnote{{The coefficients $A$ and $\mu$ are phenomenological parameters of the effective van der Waals--type fluid description. Here $A>0$ accounts for gradient effects, while $\mu=\epsilon\mu_0$ with $\epsilon\ll1$ represents a small viscosity. They encode effective dispersive and dissipative corrections.
}
},
\begin{equation}\label{u 1 dot}
\begin{split}
\dot{u}_1 =\ 
& -A s^2 x_1
+ \frac{x_1\left(8 q^3 (5 a-4 \pi  T v-5)-v^3 (a+\pi  T v-1)\right)}{\pi  v^6} \\[4pt]
& + \frac{x_1 x_2\left(v^3 (3 a+2 \pi  T v-3)-80 q^3 (3 a-2 \pi  T v-3)\right)}{2 \pi  v^7} \\[4pt]
& + \frac{\left(4 x_1^3+8 x_1 x_2^2\right)\left(240 q^3 (7 a-4 \pi  T v-7)-6 v^3 (2 a+\pi  T v-2)\right)}{32 \pi  v^8} \\[6pt]
& - \mu_0\, s\, u_1\, \varepsilon
+ \frac{\gamma \varepsilon \left(8 q^3+v^3\right)\cos (t \omega )}{v^4}
- \frac{\gamma \varepsilon\, x_2 \left(32 q^3+v^3\right)\cos (t \omega )}{2 v^5} \\[4pt]
& + \frac{\gamma \varepsilon\, x_1^2 \left(80 q^3+v^3\right)\cos (t \omega )}{2 v^6}
+ \frac{\gamma \varepsilon \left(80 q^3+v^3\right)\left(x_1^2+2 x_2^2\right)\cos (t \omega )}{4 v^6}
\end{split}
\end{equation}

\begin{equation}\label{u 2 dot}
\begin{split}
\dot{u}_2 =\ 
& -4 A s^2 x_2
+ \frac{x_2\left(8 q^3 (5 a-4 \pi  T v-5)-v^3 (a+\pi  T v-1)\right)}{\pi  v^6} \\[4pt]
& + \frac{x_1^2\left(v^3 (3 a+2 \pi  T v-3)-80 q^3 (3 a-2 \pi  T v-3)\right)}{4 \pi  v^7} \\[4pt]
& + \frac{\left(8 x_1^2 x_2+4 x_2^3\right)
\left(240 q^3 (7 a-4 \pi  T v-7)-6 v^3 (2 a+\pi  T v-2)\right)}{32 \pi  v^8} \\[6pt]
& -4 \mu_0\, s\, u_2\, \varepsilon
- \frac{\gamma \varepsilon\, x_1 \left(32 q^3+v^3\right)\cos (t \omega )}{2 v^5}
+ \frac{\gamma \varepsilon\, x_1 x_2 \left(80 q^3+v^3\right)\cos (t \omega )}{v^6}
\end{split}
\end{equation}

We linearize the differential equation by setting $\gamma = 0$ and neglecting the contributions from higher-order modes. Under this approximation, the system reduces to the following linearized form,
\begin{equation}
    \dot{Z}_{L}(t)=L {Z_{L}}(t)
\end{equation}
where $Z_{L} = (x_{1}, x_{2}, u_{1}, u_{2})^{T}$, and the linear stability matrix $L$ is given by,
\begin{equation}
L=
    \begin{bmatrix}
    0 & 0 & 1 & 0 \\
    0 & 0 & 0 & 1 \\
    -As^{2}+\Psi & 0 & -\varepsilon\mu_{0}s & 0 \\
    0 &  -4As^{2}+\Psi & 0 & -4\varepsilon\mu_{0}s
\end{bmatrix}
\end{equation}
The eigenvalues of $L$ can be written as,
\begin{equation}\label{Eigenvalues 12}
    \lambda_{L}^{1,2}=-\frac{\varepsilon\mu_{0}s}{2}\pm\frac{1}{2}\sqrt{\varepsilon^{2}\mu_{0}^{2}s^{2}-4(As^{2}-\Psi)},~
\lambda_{L}^{3,4}=-2\varepsilon\mu_{0}s\pm\sqrt{4\varepsilon^{2}\mu_{0}^{2}s^{2}-(4As^{2}-\Psi)}.
\end{equation}
Here,
\begin{equation}
\begin{split}
   \Psi = \frac{ \left(8 q^3 (5 a-4 \pi  T v-5)-v^3 (a+\pi  T v-1)\right)}{\pi  v^6}
\end{split}
\end{equation}
As evident from the eigenvalue expressions in eq.~\eqref{Eigenvalues 12}, the stability of the dynamical modes are governed by the parameter $s^2$. For $s^2 < \frac{\Psi}{A}$, the eigenvalues $\lambda_L^1 > 0$ and $\lambda_L^2 < 0$ are both real and of opposite signs. This indicates a saddle point and, consequently, an unstable equilibrium for the first mode. Conversely, when $s^2 > \frac{\Psi}{4A}$, the eigenvalues $\lambda_L^{3,4}$ become purely imaginary, signifying oscillatory behavior. In the intermediate regime $\frac{\Psi}{4A} < s^2 < \frac{\Psi}{A}$, the eigenvalues $\lambda_L^{1,2}$ retain real parts of opposite signs, with at least one being positive, further confirming the presence of a saddle point for the first node. On the other hand, both $\lambda_L^{3,4}$ possess negative real parts in this regime, indicating a stable spiral and asymptotically stable equilibrium for the second and higher modes~\cite{SLEMROD1985135}.

To proceed with analyzing the behavior of these partial differential equations under small perturbations, we recast eq.~\eqref{x dot}, eq.~\eqref{u 1 dot} and eq.~\eqref{u 2 dot} into their phase space representation from eq.~\eqref{evolution},
\begin{equation}
    \dot{Z}=f(Z)+\varepsilon g(Z,t)
\end{equation}
where, $\varepsilon \ll 1$ denotes a small perturbation parameter and $g(Z, t)$ is assumed to be a periodic function of time $t$. In the absence of perturbation, that is, for $\varepsilon = 0$, the system is reduced to the autonomous form $\dot{Z} = f(Z)$. The associated Hamiltonian then describes a smooth, energy-conserving flow, typically admitting a homoclinic orbit that connects a hyperbolic fixed point to itself. Under these conditions, the solution takes the form $Z = Z_{0}(t)$, which can be written as \cite{Zhou:2022eft},
\begin{equation}
    Z_{0}(t)=
    \begin{bmatrix}
        \frac{4c}{\sqrt{-\bar{P}_{vvv}}} \text{sech(ct)}\\
        0\\
        \frac{-4c^{2}}{\sqrt{-\bar{P}_{vvv}}} \text{sech(ct)}\tanh(ct)\\
        0\\
    \end{bmatrix}
\end{equation}
where, $c^{2}=\bar{P}_{v}-As^{2}$. The Melnikov function serves as an important mathematical tool for detecting the onset of chaotic behavior in perturbed Hamiltonian systems as already defined in eq.~\eqref{melnikov}.

\begin{figure}[ht]
  	\begin{center}
  		{\centering
  		\subfloat[]{\includegraphics[width=6cm,height=3.5cm]{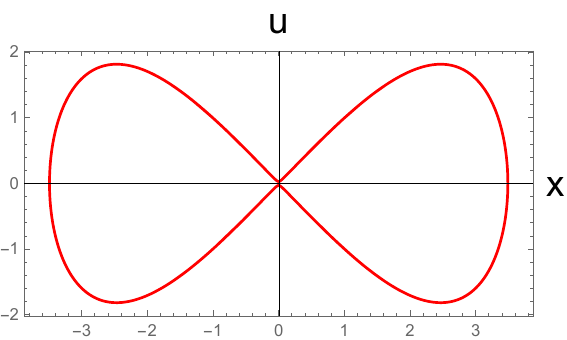} } \hspace {0.0cm}
  			\subfloat[]{\includegraphics[width=6cm,height=3.5cm]{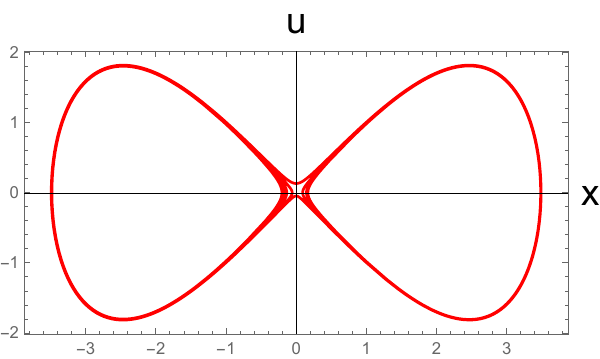} } 
  			\caption{(a) The unperturbed orbit at perturbation amplitude $\gamma=0$. (b) Phase portrait in the velocity–displacement plane for the perturbed case with perturbation amplitude $\gamma=0.002 > \gamma_c$ and perturbation parameter $\varepsilon=0.05 \neq 0$.}  \label{Temporal_1}
  		}
  	\end{center}
  \end{figure}

\begin{figure}[ht]
  	\begin{center}
  		{\centering
  		\subfloat[]{\includegraphics[width=6cm,height=3.5cm]{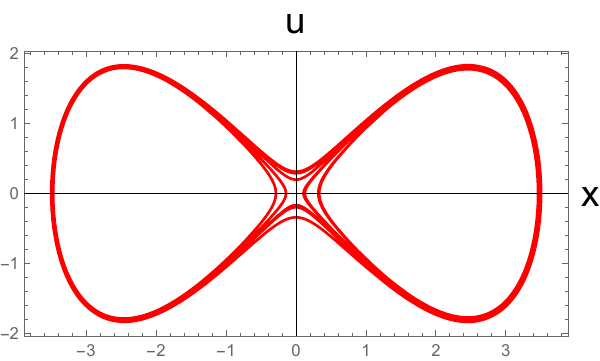} } \hspace {0.0cm}
  			\subfloat[]{\includegraphics[width=6cm,height=3.5cm]{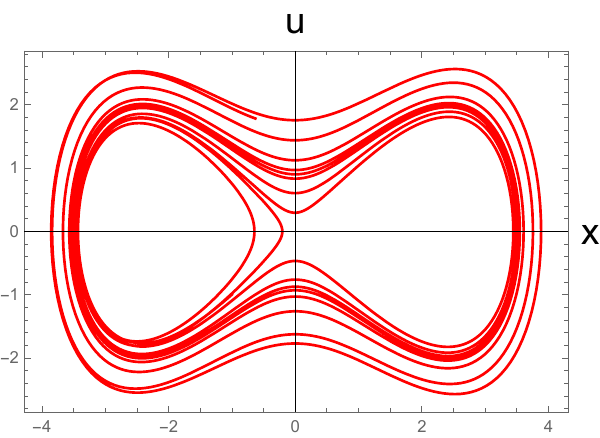} } 
  			\caption{(a) Phase portrait in the velocity–displacement plane for the perturbed case with perturbation amplitude $\gamma=0.008 > \gamma_c$ and perturbation parameter $\varepsilon=0.05 \neq 0$. (b) Phase portrait in the velocity–displacement plane for the perturbed case with perturbation amplitude $\gamma=0.04 > \gamma_c$ and perturbation parameter $\varepsilon=0.05 \neq 0$.}  \label{Temporal_2}
  		}
  	\end{center}
  \end{figure}

\begin{figure}[ht]
  	\begin{center}
  		{\centering
  		\subfloat[]{\includegraphics[width=6cm,height=3.5cm]{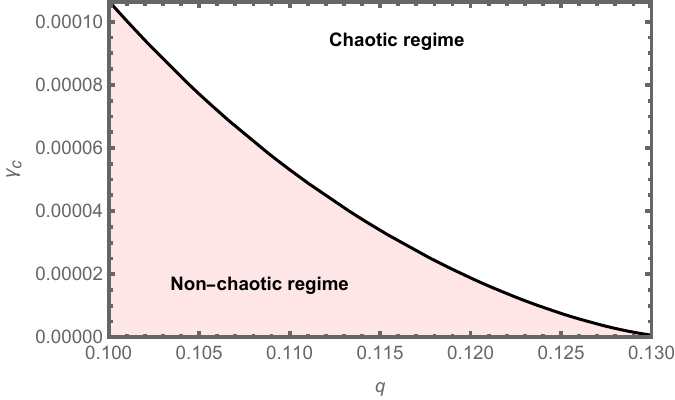} } \hspace {0.0cm}
  			\subfloat[]{\includegraphics[width=6cm,height=3.5cm]{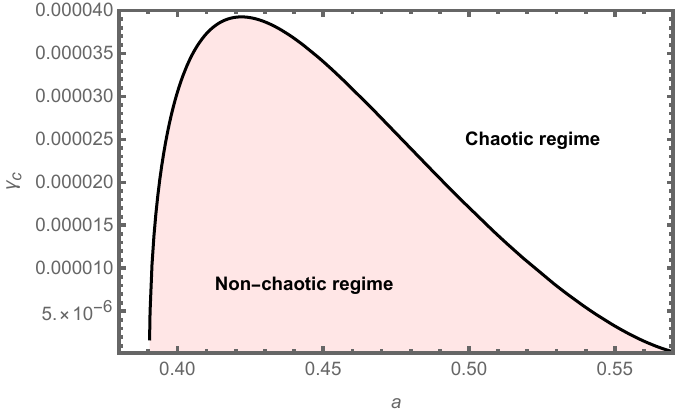}} 
  			\caption{(a) The behavior of $\gamma_c$ with charge $q$ for fixed values of $a=0.1$ and $T_0=0.1,~v=0.5,~\varepsilon=0.005, ~\mu_{0}=0.01,~A=0.05,~\omega=0.005,~s=0.001$. (b) The behavior of $\gamma_c$ with string fluid parameter $a$ for fixed values of $q=0.08$, $T_0=0.2,~v=0.5,~\varepsilon=0.005, ~\mu_{0}=0.01,~A=0.05,~\omega=0.005,~s=0.001$.}  \label{gamma c}
  		}
  	\end{center}
  \end{figure}

The phase-space separation between the stable and unstable manifolds associated with the perturbed system is quantified by the Melnikov function $M(t_0)$, which measures their transverse distance along the homoclinic (or heteroclinic) orbit. If this function possesses a basic zero for some $t_{0}$, at that point the manifolds intersect transversely, heralding chaotic dynamics via Smale-Birkhoff theorem\cite{smale1963diffeomorphisms,Bronnikov:1994ja}. The matrix $\Omega_{n}$ encodes the canonical symplectic structure of the system. When considering two degrees of freedom (i.e. a four-dimensional phase space), it explicitly takes the form,
\begin{equation}
   \Omega_{2} =  
   \begin{bmatrix}
       0 & 1 & 0 & 0 \\
       -1 & 0 & 0 & 0 \\
       0 & 0 & 0 & 1 \\
       0 & 0 & -1 & 0 \\
   \end{bmatrix},
\end{equation}
which preserves the Hamiltonian structure through Poisson bracket formulation. The existence of simple zeros in $M(t_0)$ shows that the perturbation system's steady and dynamic manifolds intersect transversely in phase space, resulting in a Smale horseshoe, which is indicative of the onset of chaos. In case of Hayward AdS black holes with string fluids, the Melnikov function can be calculated from eq.\eqref{melnikov},
\begin{equation}
\begin{split}
    M(t_{0})=\int_{-\infty}^{+\infty}&[C_{1}\sech^{2}(c(t-t_{0}))\tanh^{2}(c(t-t_{0}))+C_{2}\sech(c(t-t_{0}))\tanh(c(t-t_{0}))\\
    &(1-C_{3}\sech^{2}((c(t-t_{0})))\cos(\omega (t-t_{0}) +\omega t_{0})]dt\\
\end{split}
\end{equation}
where,
\begin{equation}
    C_{1}=\frac{16s\mu_{0}c^{4}}{\bar{P}_{vvv}},~ C_{2}=-\frac{4\gamma c^{2}\bar{P}_{T}}{\sqrt{\bar{P}_{vvv}}},~ C_{3}=\frac{6 c^{2} \bar{P}_{vvT}}{\bar{P}_{vvv}}.
\end{equation}
The explicit form of the Melnikov function can now be written as,
\begin{equation}
    M(t_{0})=\gamma \omega K \sin(\omega t_{0})+\mu_{0}sL
\end{equation}
with
\begin{eqnarray}
   K=\frac{4\pi}{v_0^{4}\sqrt{-\bar{P}_{vvv}}}\left[(8 q^3 + v_0^3)-\frac{2(80 q^3 + v_0^3)}{v_0^{2}\bar{P}_{vvv}}(\omega^{2}+c^{2})\right]\sech\left(\frac{\pi\omega}{2c}\right),~
    L=\frac{32c^{3}}{3\bar{P}_{vvv}}.
\end{eqnarray}
In order for the Melnikov function to admit a simple zero, thereby indicating the potential onset of chaos following a weak, time-periodic perturbation, it is necessary that the condition $\left|\frac{s\mu_{0}L}{\gamma \omega K}\right|\leq1$ should be satisfied. This leads to the identification of a critical parameter $\gamma_{c}$ which serves as the threshold for chaotic behavior and is given as,
\begin{equation}
    \gamma_{c}=\left| \frac{s\mu_{0}L}{\omega K}\right|
\end{equation}
For Hayward AdS black holes with string fluids, the critical threshold value of $\gamma_c$ can be computed directly as,
\begin{equation}\label{Gamma Critical}
\gamma_{c}=\frac{4 c^3 \mu _0 s v^{12} \cosh \left(\frac{\pi  \omega }{2 c}\right) \sqrt{\frac{2 v^3 (2 a+\pi  T v-2)-80 q^3 (7 a-4 \pi  T v-7)}{v^8}}}{\sigma}
\end{equation}
where,
\begin{equation}
\begin{split}
\sigma=&\sqrt{3 \pi } \omega  \Bigg(8 q^3 v^3 \left(-99 a+10 \pi  v^3 \left(c^2+\omega ^2\right)+63 \pi  T v+99\right)\\
&+v^6 \left(6 a+\pi  v^3 \left(c^2+\omega ^2\right)+3 \pi  T v-6\right)-960 q^6 (7 a-4 \pi  T v-7)\Bigg)
\end{split}
\end{equation}
In the unperturbed limit ($\varepsilon = 0$, $\gamma = 0$), where the system is free from external driving and damping, the velocity–displacement (u-x) phase portrait exhibits a symmetric, butterfly-shaped structure. This configuration arises due to the inherent nonlinearities of the system and corresponds to homoclinic orbits that connect saddle points in phase space. Such a structure reflects the presence of rich dynamical features, even in the absence of perturbative effects. The butterfly-like phase space structure discussed above is illustrated in Figure~\ref{Temporal_1}(a).
\quad Utilizing eq.~\eqref{Gamma Critical} we can compute the critical value of $\gamma$ given as, $\gamma_{c}=0.00000636301$ for fixed values of $q=0.09, a=0.4, T_{0}=0.036, v=0.48, \varepsilon=0.05,\mu_{0}=0.01,A=0.5, \omega=0.6, s=0.01$. When $\gamma > \gamma_c$ and $\varepsilon \neq 0$, the previously observed structure becomes distorted, indicating the onset of perturbative effects, as illustrated in Figure~\ref{Temporal_1}(b). Further increasing $\gamma$ beyond the critical threshold $\gamma_c$, the chaotic behavior becomes more pronounced as shown in Figure~\ref{Temporal_2}.
To study the behavior of $\gamma$ critical, we plot $\gamma_c$ with charge $q$ and the string fluid parameter $a$, as shown in Figure~\ref{gamma c}. The values of $\gamma_{c}$ required to induce chaotic behavior are illustrated there. These plots clearly delineate two distinct regimes: a chaotic regime and a non-chaotic regime. The Figure~\ref{gamma c}(a) illustrates the variance of the critical threshold $\gamma_{c}$ in relation to the charge parameter $q$, while maintaining a constant string fluid parameter $a$. The results indicate that $\gamma_{c}$ decreases as $q$ increases, suggesting that chaotic behavior becomes easier to induce at higher charge in the Hayward black hole with string fluid. The  Figure~\ref{gamma c}(b) displays the dependence of $\gamma_{c}$ on the string fluid parameter $a$, with fixed $q$. In this case, $\gamma_{c}$ initially increases rapidly and then decreases as the string fluid parameter $a$ increases, revealing a non-monotonic sensitivity of the chaotic threshold to the strength of the string fluid contribution.

 \clearpage
\section{Analysis of Spatial Chaos}\label{Spatial chaos}
Spatial chaos refers to the emergence of intricate and irregular spatial structures in a physical system, arising from intrinsic nonlinearities and dynamical instabilities. Unlike temporal chaos, which unfolds unpredictably over time, spatial chaos is characterized by complex variations across spatial dimensions, often induced by spatially periodic perturbations. In black hole thermodynamics—particularly within the extended phase space framework—spatial chaos may occur when the black hole is located in the unstable spinodal region and is subjected to modulated boundary conditions or spatially inhomogeneous external sources. For example, introducing a spatially oscillating temperature profile of the form\cite{MahishBhamidipati2019},
\begin{equation}\label{SpatialPerturbation}
    T(x) = T_0 + \varepsilon \cos(p x),
\end{equation}
where $p$ is the spatial frequency, can lead to nontrivial spatial dynamics. Such perturbations disrupt the equilibrium spatial distribution of thermodynamic quantities like the specific volume, revealing the system’s sensitivity to spatial inhomogeneities. Analytical techniques such as spatial Hamiltonian analysis and the Melnikov method provide valuable insight into the onset of spatial chaos.

In this section, we examine spatial chaos for the Hayward black hole solution in AdS with string fluids. The Hayward metric describes a class of regular black holes with no curvature singularities, owing to the presence of a de Sitter core at the origin \cite{Nascimento:2023tgw}. When immersed in a background of string fluid, characterized by anisotropic stress-energy components, the black hole exhibits modified thermodynamic properties. The incorporation of a negative cosmological constant (AdS background) facilitates a coherent thermodynamic interpretation within an extended phase space, wherein the cosmological constant is equated with pressure. To explore the effect of spatial perturbations, we model the system via an analogy with the van der Waals–Korteweg fluid, in which the mechanical stress incorporates gradient corrections and is given as~\cite{SLEMROD1985135},
\begin{equation}
\tau(x) = -\bar{P}(v(x), T) - A \frac{d^2 v}{dx^2},
\end{equation}
Here $\tau(x)$ is called Piola stress tensor and $v(x)$ is the specific volume varying in space, $T_0$ is the background temperature and $A$ is a positive parameter quantifying spatial coupling. Under equilibrium conditions ($d\tau/dx = 0$), the stress reduces to a constant ambient pressure $B$\footnote{B is the stress at $\left|x\right|=\infty$}, resulting in the nonlinear equation,
\begin{equation}\label{Spetial Final 1}
A \frac{d^2 v}{dx^2} + \bar{P}(v, T) = B.
\end{equation}
This second-order equation defines a conservative dynamical system whose solutions in the $(v, v')$ phase space may exhibit homoclinic or heteroclinic orbits\footnote{Here $v'$ is define as $\frac{dv}{dx}$ and $v''$ is define as $\frac{d^{2}v}{dx^{2}}$}, representing localized or transitionary spatial structures on the black hole horizon. From eq.~\eqref{Spetial Final 1} and the corresponding Figure~\ref{case 0}, for fixed value of $P_{0}=0.0024202$, one can identify three distinct types of orbits in the $(v'-v)$ phase plane.

\subsection*{Spatial regime: case I}\label{supercritical spatial}
\begin{figure}[ht]
    \centering
    \includegraphics[scale=.5]{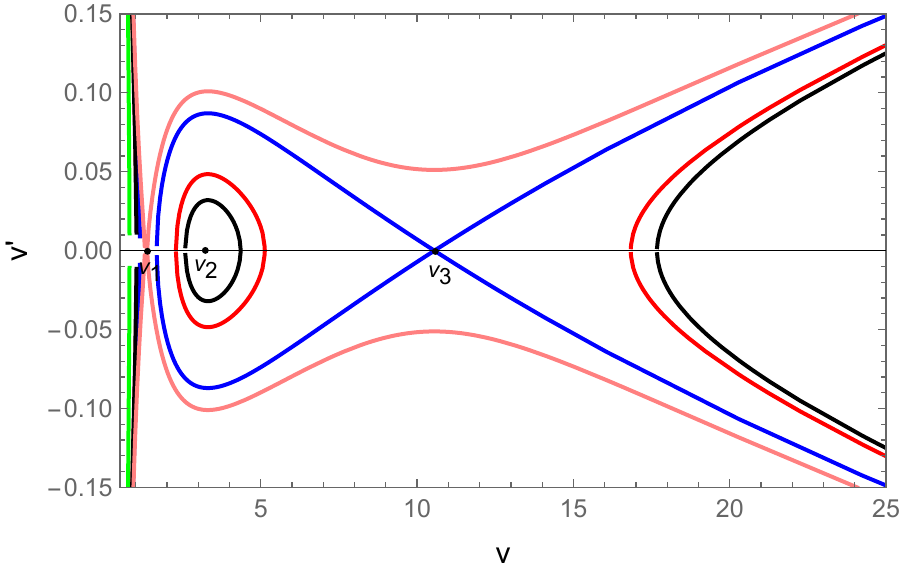}
    \vspace{5pt}
    \caption{Unperturbed phase portrait $(v'-v)$ in supercritical regime with fixed values of $q=0.4, a=0.4,T_{0}=0.036,B=0.00255629$.}
    \label{case 1}
\end{figure}

In this scenario, the reference pressure $P_{0}$ is exceeded by the ambient pressure $B$ (i.e. $B > P_0$) and the system lies on a subcritical isotherm ($T_{0} < T_{\mathrm{cr}}$), the phase portrait reveals an orbit that starts and returns to the saddle point at $v = v_3$. This closed trajectory is known as a homoclinic orbit. In Figure~\ref{case 1}, the red curve illustrates this trajectory.

\subsection*{Spatial regime: case II}\label{subcritical spatial}
\begin{figure}[ht]
    \centering
    \includegraphics[scale=.5]{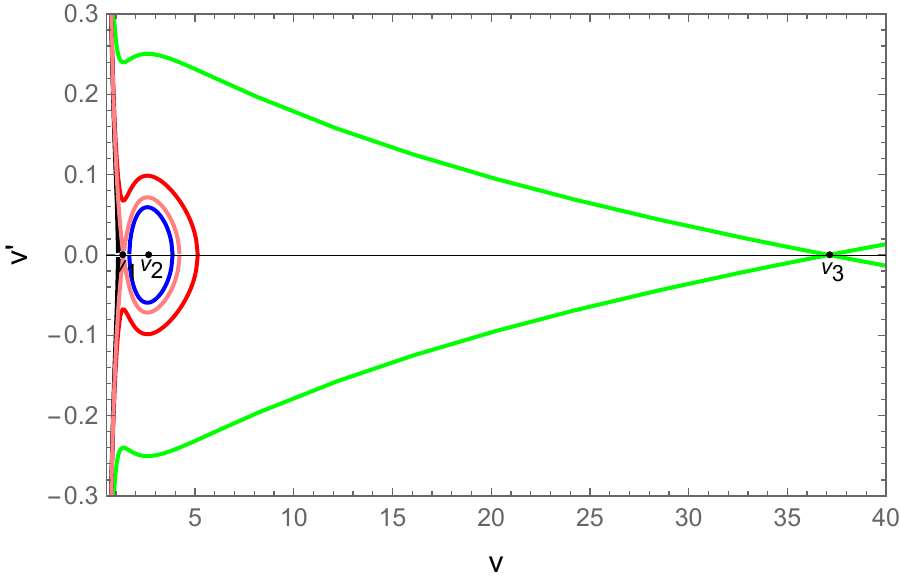} 
    \vspace{5pt}
    \caption{Unperturbed phase portrait $(v'-v)$ in subcritical regime with fixed values of $q=0.4, a=0.4,T_{0}=0.036,B=0.000901$.}
    \label{case 2}
\end{figure}

When the ambient pressure is lower than the reference pressure $P_0$ (i.e. $B < P_0$) and the system evolves along a subcritical isotherm $(T_{0} < T_{cr})$, the orbit in the $(v'-v)$ phase plane connects the saddle point at $v = v_1$ back to itself, forming a closed trajectory known as a homoclinic orbit. As shown in Figure~\ref{case 2}, the blue curve represents this trajectory.

\subsection*{Spatial regime: case III}\label{critical spatial}

\begin{figure}[ht]
    \centering
    \includegraphics[scale=.5]{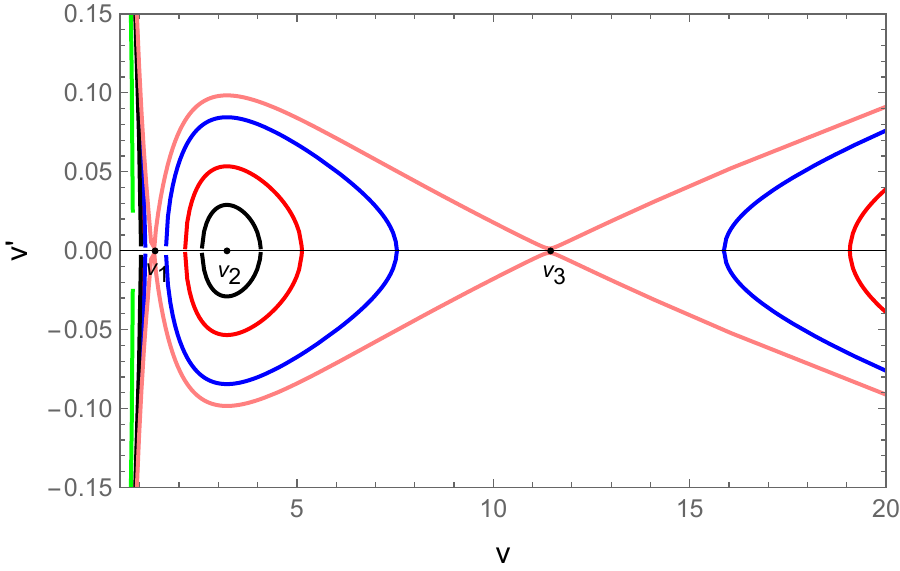} 
    \vspace{5pt}
    \caption{Unperturbed phase portrait $(v'-v)$ in critical regime with fixed values of $q=0.4, a=0.4,T_{0}=0.036,B=0.0024202$.}
    \label{Case 3}
\end{figure}

When the ambient pressure equals the reference pressure $P_0$ (i.e., $B = P_0$) and the system is considered along a subcritical isotherm ($T_{0} < T_{cr}$), the orbit in the $(v'-v)$ phase plane connects the two saddle points located at $v = v_1$ and $v = v_3$. This type of trajectory is referred to as a heteroclinic orbit. In Figure~\ref{Case 3}, the trajectory is depicted by the blue curve. In all the cases, $v_{2}$ is the center of the orbit.

To assess the impact of spatial perturbations, we substitute eq.~\eqref{SpatialPerturbation} in  the dynamical equation.

\begin{equation}\label{Spetial Final Equation}
    A \frac{d^2 v}{dx^2} + \bar{P}(v, T_0) + \frac{\varepsilon \cos(p x)}{v(x)} = B.
\end{equation}
For any subcritical temperature $T_0 < T_{\mathrm{cr}}$, the dynamical system governed by eq.~\eqref{Spetial Final Equation} admits three fixed points. These correspond to the intersection points between the isotherm $T = T_0$ and the reference pressure line $\bar{P} = B$ in the $\bar{P}$–$v$ diagram. The fixed points are ordered such that $v_1 < v_2 < v_3$. Furthermore, the second-order dynamical eq.~\eqref{Spetial Final Equation} can be recast as a system of two coupled first-order differential equations given as,
\begin{equation}
    v'=h
\end{equation}
\begin{equation}
    v''=h'=\frac{B-\bar{P}(v,T_{0})}{A}-\frac{\varepsilon\cos(px)}{v}
\end{equation}
The general forms of the solutions characterizing homoclinic and heteroclinic trajectories are obtained as,
\begin{equation}
Z(x)=
    \begin{bmatrix}
        v_{0}(x-x_{0})\\
        h_{0}(x-x_{0})\\
    \end{bmatrix}
\end{equation}
and using them in eq.~\eqref{Spetial Final Equation}, one can write,
\begin{equation}\label{f,g Spatial}
f(z_{0}(x-x_{0}))=
    \begin{bmatrix}
       h_{0}(x-x_{0})\\
       \frac{B-\bar{P}(v_{0}(x-x_{0}),T_{0})}{A})\\
    \end{bmatrix}, \qquad g(z_{0}(x-x_{0}))=\begin{bmatrix}
       0\\
      \frac{\varepsilon\cos(px)}{v_{0}(x-x_{0})}\\
    \end{bmatrix}
\end{equation}
The resulting system becomes a non-autonomous Hamiltonian system with spatial periodic forcing, making it suitable for Melnikov analysis. Melnikov function form the eq.~\eqref{melnikov} can be written for spatially perturbed system.
\begin{equation}
    M(x_{0}) = \int_{-\infty}^{+\infty} f^{T}(Z_{0}(x - x_{0})) \, \Omega_{n} \, g(Z_{0}(x - x_{0})) \, dt,  
\end{equation}
Now using eq.~\eqref{f,g Spatial} one can rewrite,
\begin{equation}
M(x_0) = \int_{-\infty}^{+\infty} \left[ -\frac{
h_{0}(x - x_0) \cos(p x)}{v_0(x - x_0)} \right] dx,
\end{equation}
where,
\begin{equation}
    \Omega_{n=1}=\begin{bmatrix}
    0 & 1 \\
    -1 & 0  \\
\end{bmatrix}
\end{equation}
where $v_0(x)$ denotes the unperturbed solution and $x_0$ is a phase shift. This function captures the separation between stable and unstable manifolds in the perturbed system and simplifies to,
\begin{equation}
    M(x_0) =  N \sin(p x_0)-L \cos(p x_0),
\end{equation}
with constants $L$ and $N$ determined by overlap integrals. A simple zero of $M(x_0)$ implies a transverse intersection of the manifolds, indicating chaotic spatial dynamics. The expression of $L$ and $N$ is given by,
\begin{equation}
   N=\int_{-\infty}^{\infty}\frac{h_{0}(x)\sin(px)}{v_{0}(x)}dx, \qquad  L=\int_{-\infty}^{\infty}\frac{h_{0}(x)\cos(px)}{v_{0}(x)}dx,
\end{equation}
In particular, the Melnikov function in this setup always possesses simple zeros for any nonzero $\varepsilon$, confirming that spatial chaos emerges even under infinitesimal perturbations. This behavior is in contrast to temporal chaos, which typically requires a finite threshold in the perturbation amplitude to manifest. As a result, the thermodynamic configuration of the black hole horizon can become unstable because of arbitrarily weak spatial modulations. Thus, spatial chaos naturally arises in regular Hayward black holes in AdS spacetime with string fluids and subjected to spatially oscillating thermal perturbations. This phenomenon demonstrates the sensitive dependence of the horizon structure on spatial inhomogeneities and reflects the intricate dynamical response of regular black holes to external fields. These results deepen our understanding of the thermodynamic phase space of regular black holes and may have broader implications for gravitational thermodynamics and holographic dualities involving smooth horizon geometries.


\section{Lyapunov Exponent as a Chaos probe}\label{Lyapunov}
Having analyzed the chaotic behavior of the system in the previous section using Melnikov's method, we now turn to the Lyapunov exponent as a complementary diagnostic of instability. Whereas the Melnikov analysis captures the onset of chaos through the global phase-space structure of the perturbed system, the Lyapunov exponent provides a local measure of the instability of circular geodesics. In this way, it allows us to examine how the black hole phase structure is reflected in the corresponding geodesic dynamics.

\subsection*{Lyapunov exponent from linearized flow}

Consider a dynamical system on phase space with coordinates $X_i(t)$,
\begin{equation}
\frac{dX_i}{dt}=\mathcal{A}_i(X)\,.
\end{equation}
Where $\mathcal{A}_i(X)$ denote the components of the dynamical flow in phase space. Let $X_i(t)$ be a reference solution (e.g.\ a circular orbit) and perturb it as
$X_i\to X_i+\delta X_i$. On linearizing it, we get;
\begin{equation}
\frac{d}{dt}\,\delta X_i(t)=K_{ij}(t)\,\delta X_j(t),\qquad 
K_{ij}(t)=\left.\frac{\partial \mathcal{A}_i}{\partial X_j}\right|_{X(t)}.
\end{equation}
Now, introducing the fundamental matrix $L_{ij}(t)$ via $\delta X_i(t)=L_{ij}(t)\,\delta X_j(0)$,
one have
\begin{equation}
\dot L_{ij}(t)=K_{im}(t)\,L_{mj}(t),\qquad L_{ij}(0)=\delta_{ij}.
\end{equation}
The (principal) Lyapunov exponent is then defined by the long-time growth rate of the linearized evolution,
\begin{equation}
\lambda=\lim_{t\to\infty}\frac{1}{t}\ln\left(\frac{L_{jj}(t)}{L_{jj}(0)}\right),
\end{equation}

\subsection*{Specialization to radial perturbations about a circular geodesic}

For circular geodesics, the relevant stability problem reduces to the two-dimensional subspace spanned by the radial degree of freedom and its conjugate momentum. 
Equivalently, one may work with the pair $(r,\dot r)$ (or $(r,p_r)$), and track an infinitesimal radial displacement $\delta r(t)$.
In this reduced sector the linear stability matrix can always be written in the canonical off-diagonal form
\begin{equation}
K_{ij}=
\begin{pmatrix}
0 & K_1\\
K_2 & 0
\end{pmatrix},
\end{equation}
so its eigenvalues are $\pm\sqrt{K_1K_2}$. One may express these entries directly in terms of the geodesic Lagrangian $\mathcal{L}$ and the background metric component $g_{rr}$ as
\begin{equation}
K_1=\frac{d}{dr}\left(\dot t^{-1}\,\frac{\delta \mathcal{L}}{\delta r}\right),\qquad 
K_2=-(\dot t\,g_{rr})^{-1},
\end{equation}
All quantities are evaluated at the circular orbit. The instability rate measured with respect to the coordinate time $t$ is therefore given by $\lambda=\sqrt{K_1K_2}$, where $\lambda>0$ corresponds to an unstable circular orbit, while $\lambda=0$ signals marginal stability. For circular geodesics, the conditions $V_r(r)=V_r'(r)=0$ hold, and the Lyapunov exponent takes the general form~\cite{Guo:2022kio,Cardoso:2008bp,Gogoi:2024akv}
\begin{equation}\label{Lyapunov General Form}
\lambda=\sqrt{-\frac{V_r''}{2\dot{t}^{\,2}}} .
\end{equation}

\subsection{Circular Geodesics and Radial Stability}

\begin{figure}[ht]
    \centering
    \includegraphics[scale=.5]{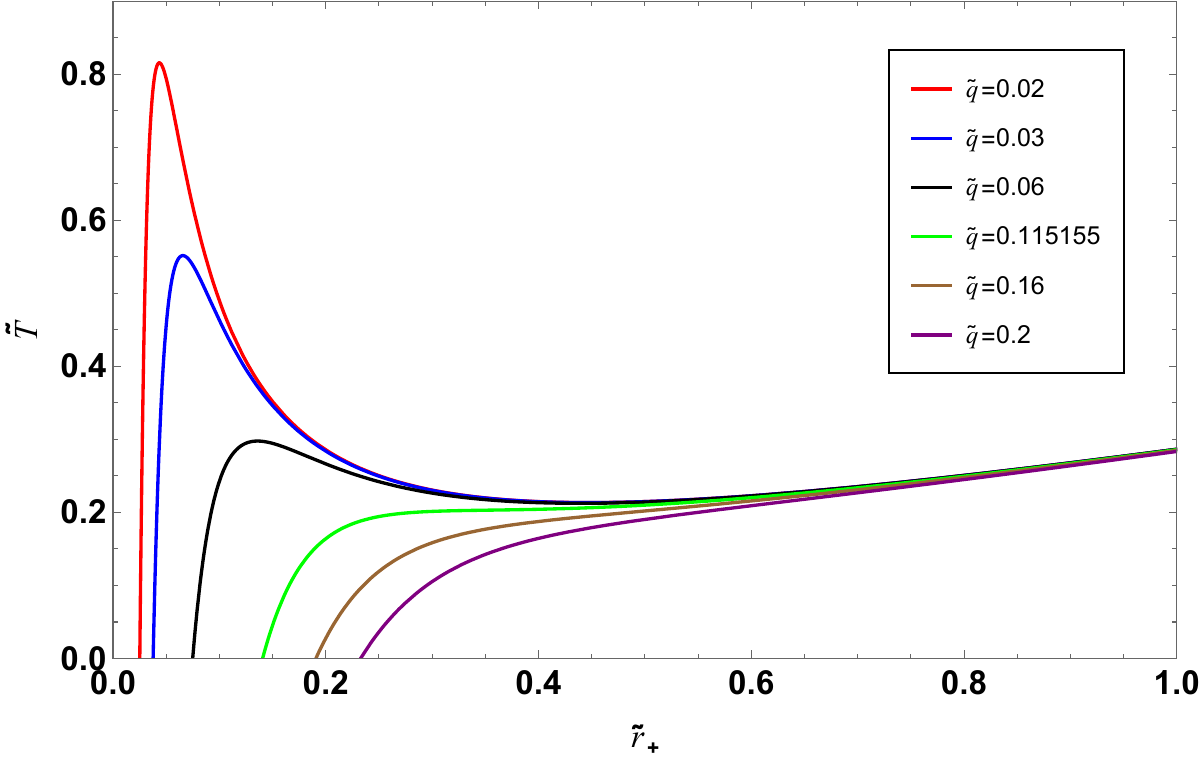} 
    \vspace{5pt}
    \caption{Hawking temperature plotted against the horizon radius for several values of $\tilde q$. At critical value $\tilde q_c=0.115155$, with the critical curve shown in green.}
    \label{Tempvsrp}
\end{figure}

\begin{figure}[ht]
    \centering
    \includegraphics[scale=.5]{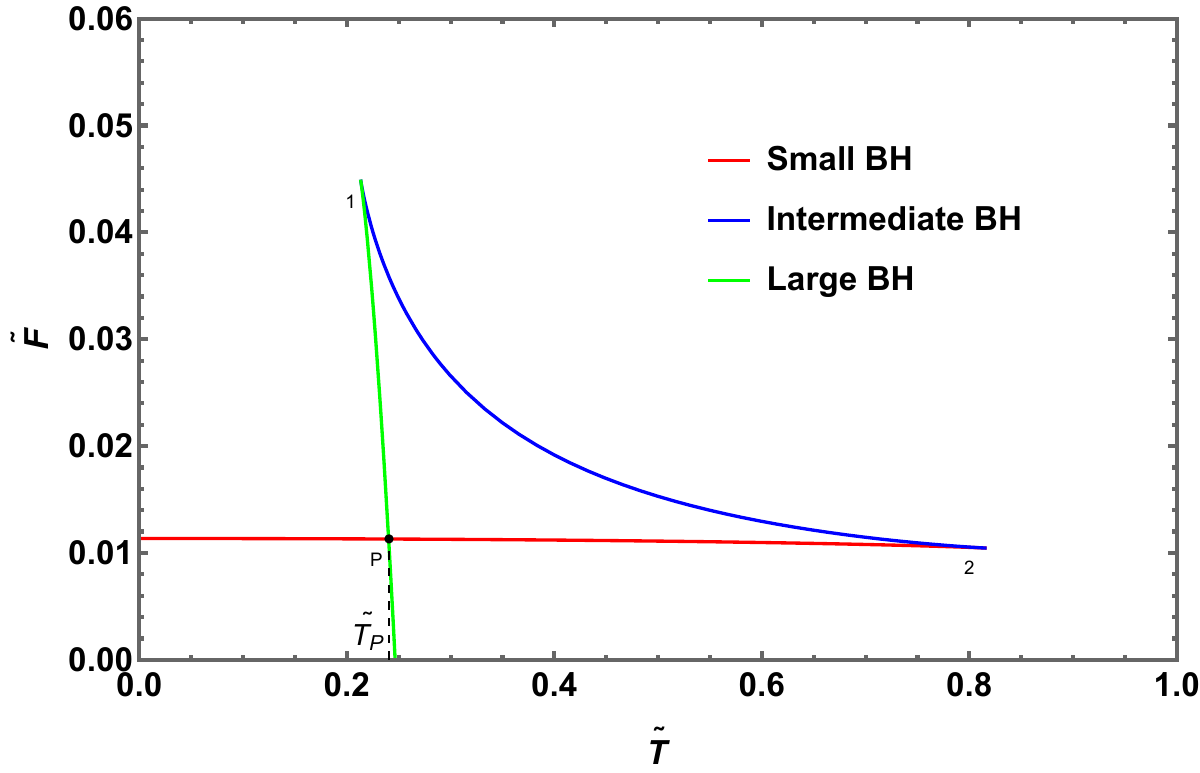} 
    \vspace{5pt}
    \caption{Three black hole phases appearing in the Gibbs free energy versus temperature diagram for $\tilde q=0.02<\tilde q_c$ at $a=0.4$.}
    \label{Gibbs}
\end{figure}

The dynamical structure and stability characteristics of black hole spacetimes can be naturally probed using circular geodesics. The regular core and the existence of a string fluid in the Hayward–AdS geometry cause nontrivial changes to the effective potential that controls geodesic motion. The presence and stability of circular orbits can be drastically changed by these factors, especially for null paths. In this subsection, we construct the effective radial potential and determine the relevant equilibrium conditions in order to examine circular geodesics and their radial stability for the Hayward–AdS black hole surrounded by a string fluid. The general form of the metric is given in eq.\eqref{General Metric}.
Where \(f(r)\) is defined in Eq.~\eqref{fr}, and the corresponding temperature is obtained from eq.~\eqref{temperature}. The Lagrangian is given by,
\begin{equation}
    2\mathcal{L}=-f(r)\dot{t}^2+\frac{\dot{r}^2}{f(r)}+r^2 \dot{\Omega}^2
\end{equation}
To find the generalized momentum $(\mathcal{P})$ we introduce,
$\mathcal{P}_q={\partial \mathcal{L}}/{\partial \dot{Q}}$
, where $Q$ is the generalized co-ordinate. We can derive the canonical momenta for each coordinate
\begin{eqnarray}
    \mathcal{P}_t&=& -\left(1 - a - \frac{2 M r^2}{q^3 + r^3} +\frac{r^2}{l^2}\right)\dot{t}=-E, \qquad \dot{t}=\frac{E}{\left(1 - a - \frac{2 M r^2}{q^3 + r^3} +\frac{r^2}{l^2}\right)}\nonumber \\ 
    \mathcal{P}_r&=& \frac{\dot{r}}{ \left(1 - a - \frac{2 M r^2}{q^3 + r^3} +\frac{r^2}{l^2}\right)}, \qquad   \mathcal{P}_{\Omega}=r^2 \dot{\Omega}=L,\qquad \dot{\Omega}=\frac{L}{r^2}
\end{eqnarray}
The Hamiltonian of the system can be expressed as,
\begin{equation}
\begin{split}
    2\mathcal{H}&=-f(r)\dot{t}^2+\frac{1}{f(r)}\dot{r}^2+r^2 \dot{\Omega}^2=-\Delta
\end{split}
\end{equation}
Further, the general form of the potential is given as, 
\begin{equation}\label{General Form Of Potential}
    V_r=f(r)\left[\Delta-\frac{E^2}{f(r)}+\frac{L^2}{r^2}\right]
\end{equation}
For timelike geodesics, one can set \(\Delta = 1\), while for null geodesics \(\Delta = 0\). On performing a dimensional analysis and rescaling the following quantities as follows:
 \begin{equation}
     \tilde{r}_+=\frac{r_+}{l},~ \tilde{q}=\frac{q}{l},~\tilde{M}=\frac{M}{l}~~\text{and}~ \tilde{T}=Tl,
 \end{equation}
 In terms of this substitution, the dimensionless form of mass and hawking temperature,
\begin{equation}\label{Dimensionaless T}
    \tilde{M}=\frac{(1 - a + \tilde{r}_+^2) (\tilde{q}^3 + \tilde{r}_+^3)}{2\tilde{r}_+^2},\qquad\tilde{T}=\frac{2 (a-1) \tilde{q}^3+\tilde{r}_+^3 \left(-a+3 \tilde{r}_+^2+1\right)}{4 \pi  \tilde{r}_+ \left(\tilde{q}^3+\tilde{r}_+^3\right)}
\end{equation}
And the dimensionless entropy can be computed following exactly same analysis.
\begin{equation}
    \tilde{S}=2\pi\left(\frac{\tilde{r}_+^2}{2}-\frac{\tilde{q}^3}{\tilde{r}_+}\right)
\end{equation}
The Gibbs free energy is computed and expressed as below,
\begin{equation}\label{Gibbs Free Energy}
    \tilde{F}=\tilde{M}-\tilde{T}\tilde{S}=\frac{2 \tilde{q}^6 (-1 + a + \tilde{r}_+^2) - \tilde{r}_+^6 (-1 + a + \tilde{r}_+^2) + 
 2 \tilde{q}^3 \tilde{r}_+^3 (4 - 4 a + 5 \tilde{r}_+^2)}{4 \tilde{r}_+^2 (\tilde{q}^3 + \tilde{r}_+^3)}
\end{equation}
It is straightforward to evaluate the critical quantities applying the condition,
\begin{equation}
    \frac{\partial \tilde{T}}{\partial \tilde{r}_+}=\frac{\partial^2 \tilde{T}}{\partial \tilde{r}_+^2}=0
\end{equation}
 The corresponding value of critical quantities are expressed as,
 \begin{equation}
     \tilde{r}_{+c}=0.435773\sqrt{1-a},~\tilde{q}_{c}=0.176261(1-a)^{\frac{5}{6}},~\tilde{T}_c=0.264695\sqrt{1-a}
 \end{equation}
Employing the dimensionless Hawking temperature in Eq.~(\ref{Dimensionaless T}), we plot the temperature against the horizon radius for various values of \(\tilde q\), as shown in Figure.~\ref{Tempvsrp}. It is evident that for \(\tilde q < \tilde q_c\), multiple black hole configurations can occur, characterized by distinct horizon radii \(\tilde r_+\). However, once \(\tilde q\) exceeds the critical value \(\tilde q_c\), the system admits only a single black hole configuration throughout the corresponding range of horizon radius and temperature.
The phase structure can be inferred from the Gibbs free energy in Figure \ref{Gibbs}. Below the critical charge, $\tilde q_c$, the free-energy diagram develops three distinct branches, representing the small, intermediate, and large black hole solutions. These branches are present only in the temperature range $\tilde T_1<\tilde T<\tilde T_2$, where $\tilde T_1$ and $\tilde T_2$ correspond to the temperatures at points $1$ and $2$. The point $p$, located at $\tilde T_p=0.240247$, signals the phase transition temperature. To obtain these curves, we start from the Gibbs free energy in eq.~\eqref{Gibbs Free Energy} and use eq.~\eqref{Dimensionaless T} to rewrite the horizon radius $\tilde r_+$ as a function of the Hawking temperature $\tilde T$. Because the relation $\tilde r_+(\tilde T)$ is multivalued, the resulting rescaled free energy $\tilde F$ also becomes multivalued when expressed in terms of $\tilde T$ and $\tilde q$.

\subsection{Timelike Geodesics}\label{Sec. Timelike Geodesics}
\begin{figure}[ht]
    \centering
    \includegraphics[scale=0.5]{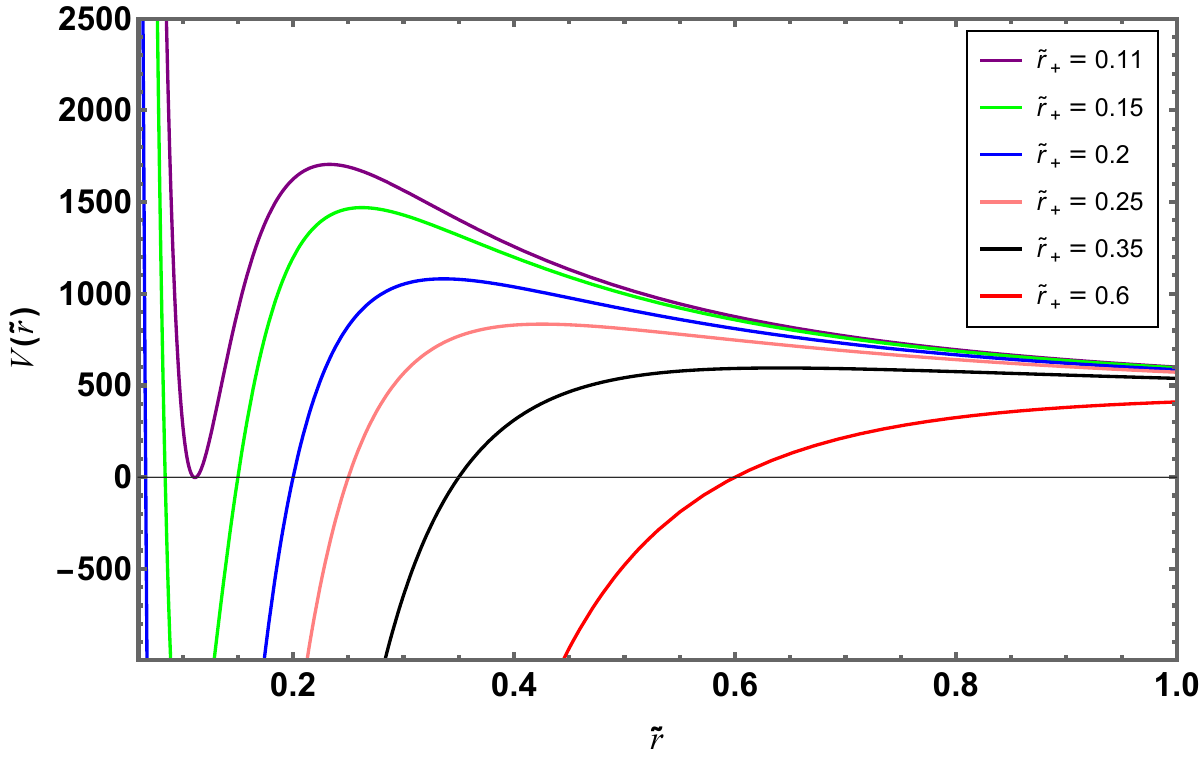} 
    \vspace{5pt}
    \caption{The effective potential $V(\tilde{r})$ is plotted as a function of $\tilde{r}$ for $\tilde{q}=0.09$ and $a=0.4$ with different black hole radius $\tilde{r}_+$. }
    \label{Potential vs r}
\end{figure}

The effective potential for the timelike case is given by eq.~\eqref{General Form Of Potential},
\begin{equation}
    V(r)=f(r)\left[1-\frac{E^2}{f(r)}+\frac{L^2}{r^2}\right]
\end{equation}
For circular orbits $V(r)=V^{'}(r)=0$
\begin{equation}
    E^2 =\frac{2f(r_0)^2}{2f(r_0)-r_{0}f'(r_0)}, \qquad L^2=\frac{r_0^3 f'(r_0)}{2f(r_0)-r_0f'(r_0)}
\end{equation}
To get real energy condition,
\begin{equation}
    2f(r_0)-r_0f'(r_0)>0,
\end{equation}
Figure~\ref{Potential vs r} shows $V_r$ as a function of $\tilde r$ for different values of $\tilde r_+$ at fixed $\tilde q=0.09$, with $L=20l$ and $E=0$. As $\tilde r_+$ increases, the maximum of the effective potential decreases and disappears for $\tilde r_+=0.6$, indicating that unstable timelike geodesics are absent for sufficiently large horizon radius.\\
The expression of $\dot{t}$ at circular geodesics can be written as,
\begin{equation}\label{Timelike t dot}
    \dot{t}=\frac{1}{\sqrt{f(r_0)-\frac{1}{2}r_0 f^{'}(r_0)}},
\end{equation}
The behavior of the effective potential for circular geodesics as a function of $\tilde r$, evaluated at $\tilde r_0$.

\begin{equation}
    V^{'}(\tilde{r}_0)=\frac{\left(\begin{split}
    &2 \bigg\{\tilde{L}^2 \left(\tilde{r}_0^5 ((a-1) \tilde{r}_0+3 \tilde{M})+(a-1) \tilde{q}^6+2 (a-1) \tilde{q}^3 \tilde{r}_0^3\right)\\
    &\qquad\qquad\qquad\qquad+ \tilde{r}_0^4 \left(\tilde{M} \left(\tilde{r}_0^3-2 \tilde{q}^3\right)+\left(\tilde{q}^3+\tilde{r}_0^3\right)^2\right)\bigg\}
    \end{split}\right)}{ \tilde{r}_0^3 \left(\tilde{q}^3+\tilde{r}_0^3\right)^2},
\end{equation}
The second radial derivative of the effective potential evaluated at the circular radius
\begin{equation}\label{Double derivative of time like V}
     V^{''}(\tilde{r}_0)=\frac{2\Pi}{ \tilde{r}_0^4 \left(\tilde{q}^3+\tilde{r}_0^3\right)^3},
\end{equation}
The Lyapunov exponent associated with timelike circular geodesics is obtained from Eq.~\eqref{Lyapunov General Form} by using Eqs.~\eqref{Timelike t dot} and \eqref{Double derivative of time like V}.

\begin{equation} \label{Timelike Lyapunov Final Form}
\lambda_{T}
=\sqrt{
\frac{\Theta\Pi
}{
\tilde r^{\,4}_0\left(\tilde q^{\,3}+\tilde r^{\,3}_0\right)^{5}
}
}\,.
\end{equation}
Where the explicit form of $\Theta$ and $\Pi$ is
\begin{equation}
     \Theta=\left(\tilde r_0^{5}\big((a-1)\tilde{r}_0+3\tilde M\big)
+(a-1)\tilde q^{\,6}
+2(a-1)\tilde q^{\,3}\tilde r^{3}_0\right),
\end{equation}

\begin{equation}
\begin{split}
\Pi&=3\tilde{L}^{2}\left\{\tilde q^{3}\tilde r^{5}_0\big(2\tilde M-3(a-1)\tilde{r}_0\big)
-\tilde r^{8}_0\big((a-1)\tilde{r}_0+4\tilde M\big)
-(a-1)\tilde q^{9}
-3(a-1)\tilde q^{6}\tilde r^{3}_0\right\}\\
&\qquad+\tilde r^{4}_0\left\{\left(\tilde q^{3}+\tilde r^{3}_0\right)^{3}
-2\tilde M\left(\tilde q^{\,6}-7\tilde q^{\,3}\tilde r^{\,3}_0+\tilde r^{\,6}_0\right)\right\},
\end{split}
\end{equation}

\begin{figure}[ht]
  	\begin{center}
  		{\centering
  		\subfloat[]{\includegraphics[width=7cm,height=5.5cm]{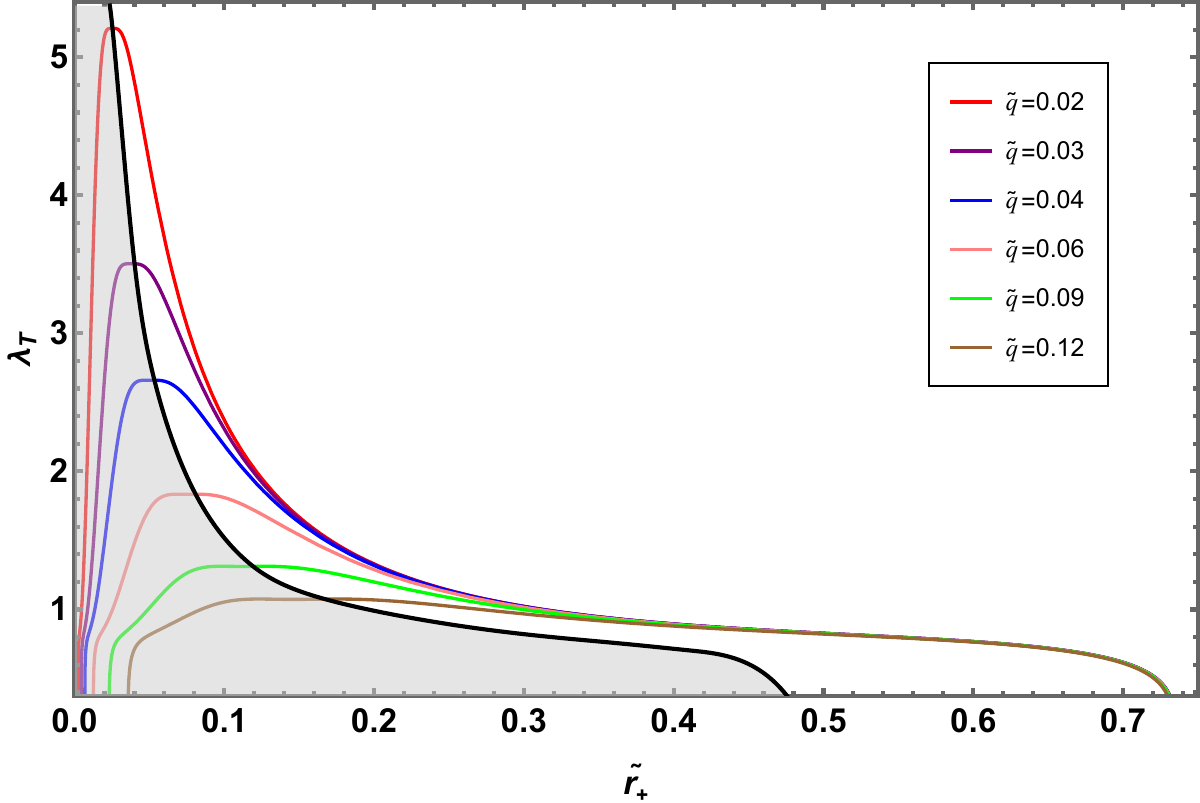} } \hspace {0.0cm}
  			\subfloat[]{\includegraphics[width=7cm,height=5.5cm]{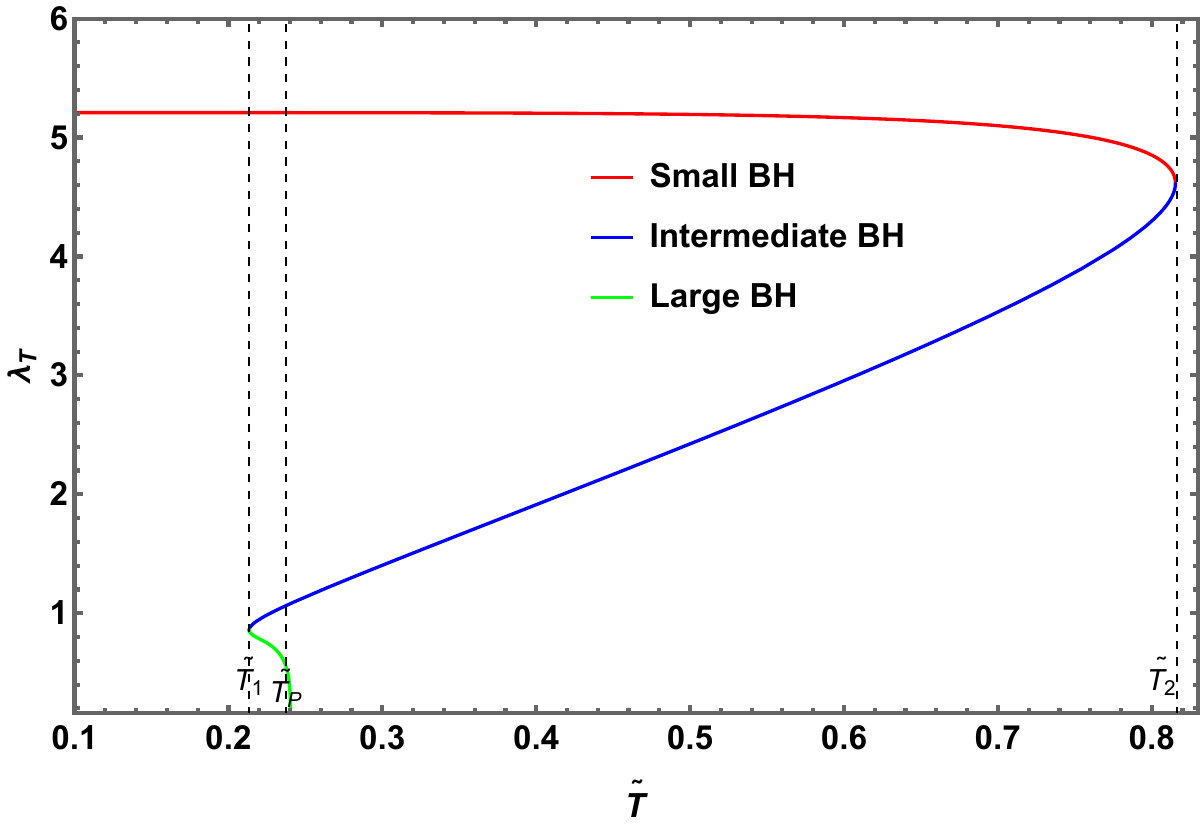} } 
  			\caption{(a) Lyapunov exponent curves for unstable circular geodesics corresponding to different values of $\tilde q$ and $a=0.4$.. (b) Lyapunov exponent versus temperature $\tilde T$ for unstable circular timelike geodesics at $\tilde{q}=0.02$ and $a=0.4$.}
  		\label{Timelike Lambda_vs_rplus_temp}}
        \end{center}
  \end{figure}

The final expression for the timelike Lyapunov exponent $\lambda_T$ is given in Eq.~\eqref{Timelike Lyapunov Final Form}. As is clear from this expression, $\lambda_T$ depends on $\tilde{r}_0$, $\tilde{M}$, $\tilde{q}$, $\tilde{L}$, and $a$, where $\tilde{L}=L/l$. For a more convenient analysis, we express $\tilde{M}$ and $\tilde{r}_0$ in terms of the horizon radius $\tilde{r}_+$. In addition, by fixing $\tilde{L}=20$ and $a=0.4$, the timelike Lyapunov exponent becomes a function only of $\tilde{r}_+$ and $\tilde{q}$. The resulting behavior of $\lambda_T$ as a function of $\tilde{r}_+$ for several values of $\tilde{q}$ is shown in Figure~\ref{Timelike Lambda_vs_rplus_temp}(a). The behavior of $\lambda_T$  depends sensitively on the value of $\tilde q$, indicating that the instability of timelike circular geodesics is strongly influenced by the black hole charge. In particular, $\lambda_T$ decreases as the charge parameter is increased. The gray-shaded region corresponds to the unphysical domain, since the associated Hawking temperature becomes negative there, while the black curve marks the limiting configuration for which the Hawking temperature vanishes. It is further observed that, for nearly all values of $\tilde q$, the timelike Lyapunov exponent drops to zero at a particular value of the horizon radius. This suggests that, for sufficiently large black holes, unstable circular timelike orbits cease to exist, so that massive particles no longer admit an unstable equilibrium configuration in this regime. 
Finally, by expressing $\tilde{r}_+$ in terms of the temperature $\tilde{T}$, we study the behavior of the timelike Lyapunov exponent $\lambda_T$ as a function of $\tilde{T}$ for $\tilde{q}=0.02<\tilde{q}_c$. The corresponding profile is shown in Fig.~\ref{Timelike Lambda_vs_rplus_temp}(b). In the temperature interval $\tilde{T}_1<\tilde{T}<\tilde{T}_2$, three branches of Hayward-AdS black holes appear, corresponding to the small, intermediate, and large black hole phases, respectively. This indicates that, within this temperature range, the system admits multiple competing black hole branches. At the phase transition temperature $\tilde{T}_P$, a first-order phase transition occurs. It is also observed from the figure that the timelike Lyapunov exponent $\lambda_T$ for the massive particle vanishes at a finite temperature, indicating that the corresponding timelike circular geodesic becomes marginally stable.


\subsection{Null Geodesics}
\begin{figure}[ht]
  	\begin{center}
  		{\centering
  		\subfloat[]{\includegraphics[width=7cm,height=5.5cm]{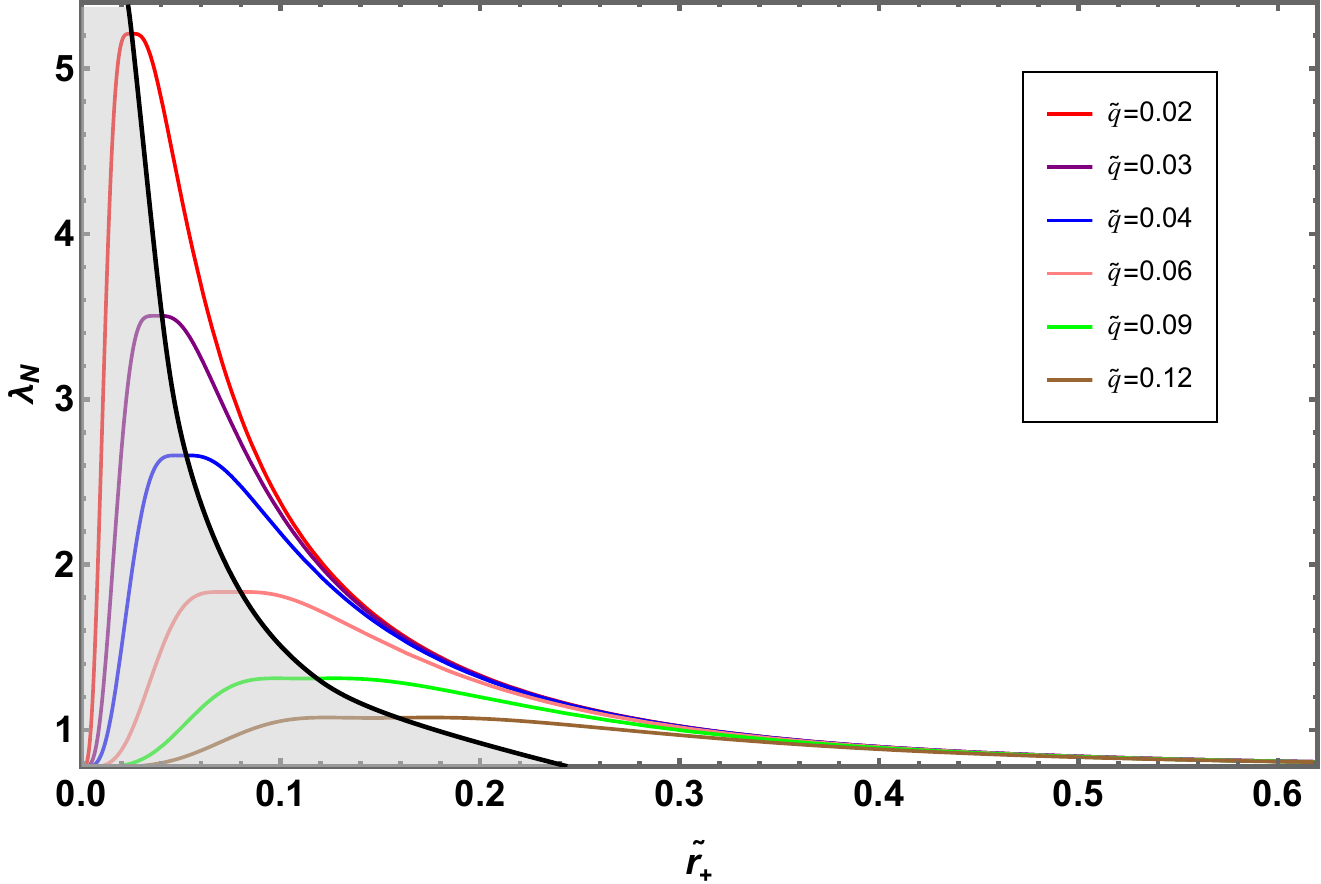} } \hspace {0.0cm}
  			\subfloat[]{\includegraphics[width=7cm,height=5.5cm]{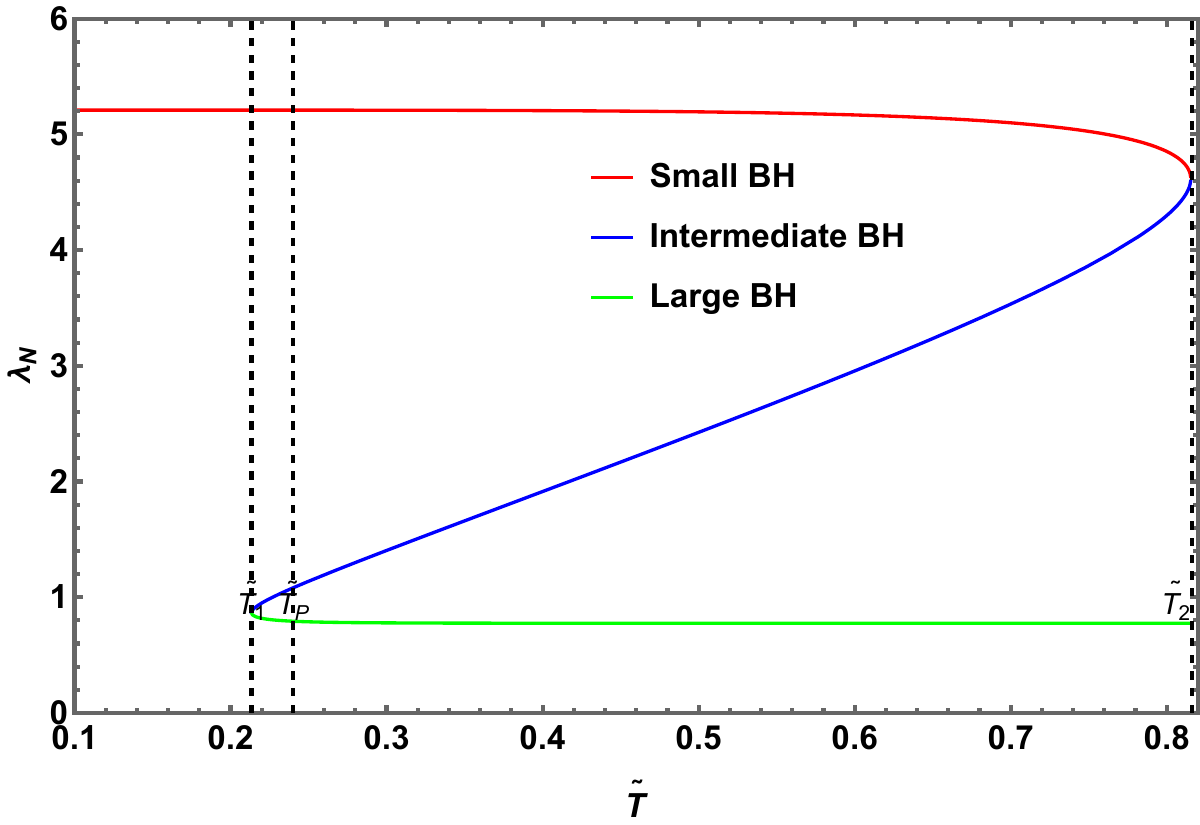} } 
  			\caption{(a) Lyapunov exponent curves for unstable circular null geodesics corresponding to different values of $\tilde q$ and $a=0.4$. (b) Null Lyapunov exponent as a function of temperature $\tilde T$ for unstable circular null geodesics at $\tilde{q}=0.02$ and $a=0.4$.}
  		\label{LamdaNull_vs_rPlus_temp}}
        \end{center}
  \end{figure}

Following \cite{Cardoso:2008bp}, the potential for null geodesics,
\begin{equation}
    V(r)=f(r)\left[-\frac{E^2}{f(r)}+\frac{L^2}{r^2}\right]
\end{equation}
Condition for unstable geodesics is $V(r_0)=V^{'}(r_0)=0$ and $V^{''}(r_0)<0$. In this condition the relation between $E$ and $L$ and the expression $\dot{t}$
\begin{equation}
    \frac{E}{L}=\frac{\sqrt{f(r_0)}}{r_0},\qquad
    \dot{t}=\frac{L}{r_0 \sqrt{f(r_0)}}
\end{equation}
Similarly to the previous section, the first radial derivative of the effective potential for null geodesics is given by the condition that governs the photon-sphere radius, where the radial potential reaches an extremum.
\begin{equation}
   V^{'}(\tilde{r}_0)=- \frac{2 \tilde{L}^2 \left\{\tilde{r}_0^5 (-(a-1) \tilde{r}_0-3 \tilde{M})-(a-1) \tilde{q}^6-2 (a-1) \tilde{q}^3 \tilde{r}_0^3\right\}}{\tilde{r}_0^3 \left(\tilde{q
   }^3+\tilde{r}_0^3\right)^2}
\end{equation}
The second radial derivative of the effective potential for null geodesics provides information about the stability of the photon sphere. A positive second derivative indicates a local minimum (stable orbit), while a negative second derivative corresponds to a local maximum (unstable orbit).
\begin{equation}
    V^{''}(\tilde{r}_0)=\frac{6 \tilde{L}^2 \Phi}{\tilde{r}_0^4 \left(\tilde{q}^3+\tilde{r}_0^3\right)^3}
\end{equation}
The Lyapunov exponent for null geodesics is determined by the second radial derivative of the effective potential. It quantifies the rate of separation between infinitesimally close trajectories near the photon sphere. Specifically, the Lyapunov exponent for null circular orbits is given by:
\begin{equation}
\lambda_N
=\sqrt{3}\,
\left(
\frac{\Phi\Gamma
}{
\tilde r_0^{\,2}\big(\tilde q^{\,3}+\tilde r_0^{\,3}\big)^{4}
}
\right)^{\frac{1}{2}}
\end{equation}
Where $\Phi$ and $\Gamma$ can be written explicitly as,
\begin{equation}
    \Phi=\left\{\tilde q^{\,3}\tilde r_0^{\,5}\big(2\tilde M-3(a-1)\tilde r_0\big)
-\tilde r_0^{\,8}\big((a-1)\tilde r_0+4\tilde M\big)
-(a-1)\tilde q^{\,9}
-3(a-1)\tilde q^{\,6}\tilde r_0^{\,3}\right\},
\end{equation}
\begin{equation}
    \Gamma=\left\{\big(a-\tilde r_0^{\,2}-1\big)\big(\tilde q^{\,3}+\tilde r_0^{\,3}\big)
+2\tilde M\,\tilde r_0^{\,2}\right\},
\end{equation}
The discussion for null circular geodesics closely parallels that of the timelike case in Section \ref{Sec. Timelike Geodesics} and will therefore be kept brief. Using the final expression for the null Lyapunov exponent $\lambda_N$, we rewrite $\tilde{M}$ and $\tilde{r}_0$ in terms of the horizon radius $\tilde{r}_+$. After fixing the remaining parameters, $\lambda_N$ depends only on $\tilde{r}_+$ and $\tilde{q}$. In Figure~\ref{LamdaNull_vs_rPlus_temp}(a), we display the corresponding behavior of $\lambda_N$ as a function of $\tilde{r}_+$ for different values of $\tilde{q}$. Figure shows that the null Lyapunov exponent $\lambda_N$ is sensitive to the charge parameter $\tilde q$, decreasing as $\tilde q$ is increased. The gray region denotes the unphysical branch, where the Hawking temperature is negative, whereas the black curve marks the boundary defined by vanishing Hawking temperature. In contrast to the timelike case, $\lambda_N$ remains nonzero in the large-black-hole regime and, for sufficiently large $\tilde r_+$, the curves for different values of $\tilde q$ approach the same asymptotic value. This indicates that, in the large-radius regime, the effect of the charge on the instability of circular null geodesics becomes subdominant, so that the orbital instability is controlled primarily by the asymptotic black hole geometry. 
Finally, we study the temperature dependence of the null Lyapunov exponent $\lambda_N$ by expressing $\tilde{r}_+$ in terms of $\tilde{T}$ for $\tilde{q}=0.02<\tilde{q}_c$, as shown in Figure~\ref{LamdaNull_vs_rPlus_temp}(b). As in the timelike case, the interval $\tilde{T}_1<\tilde{T}<\tilde{T}_2$ contains three branches of Hayward-AdS black holes, namely the small, intermediate, and large black hole phases. Thus, within this temperature window, the system admits multiple competing branches and undergoes a first-order phase transition at $\tilde{T}_P$. In contrast to the timelike case, however, the null Lyapunov exponent $\lambda_N$ does not vanish at any finite temperature.

\section{Remarks}\label{Remarks}
In the present work, we investigate the onset of chaos in the extended thermodynamic phase space of Hayward AdS black holes with string fluids, with particular attention to dynamical instabilities in the spinodal region pertaining to the phase transition across small and large black holes. We examine the perturbed Hamiltonian system derived from the thermodynamic equation of state of the black hole, taking into account temporal and spatial periodic perturbations. The system displays nonlinearities that can result in both homoclinic and heteroclinic orbits in phase space. When Melnikov's approach is used, chaotic behavior can be identified by computing Melnikov function zeros, which indicates that there are transverse intersections in the perturbed manifold, signaling the onset of chaos.

\par To construct the Hamiltonian framework corresponding to the effective fluid dynamics of the Hayward AdS black hole with string fluids, we formulated the dynamical equations governing the extended thermodynamic phase space and incorporated a weak time-periodic perturbation. Figure~\ref{Temporal_1}(a) provides the analytical identification of the homoclinic orbit in the unperturbed system employing the Melnikov method. When the perturbation amplitude exceeds a critical threshold $\gamma_c$, the system experiences a qualitative change where the homoclinic structure breaks down, resulting in temporal chaos, which is characterized by an evolution of sensitive and irregular phase space trajectories. This transition is evident, as demonstrated in Figures~\ref{Temporal_1}(b), and Figure~\ref{Temporal_2}, which exhibit the onset of chaotic behavior. This phenomena validates the claim that chaotic instabilities under time-periodic perturbations are generically supported by nonlinear thermodynamic dynamics in the spinodal region, thus being consistent with prior assessments of chaos in charged AdS black hole structures~\cite{Zhou:2022eft,Kumar:2024qon,MahishBhamidipati2019,SLEMROD1985135,Chabab:2018lzf,Barzi:2024bbj}. Additionally, we investigated how the intrinsic parameters of the Hayward AdS black hole with string fluids, notably the charge $q$ and the string fluid coupling $a$, affects the critical perturbation amplitude $\gamma_c$. Eq.~\eqref{Gamma Critical} contains the analytical expression for $\gamma_c$, which highlights how regularization parameters and nonlinear matter content influence the system's sensitivity to chaos, demonstrated graphically in Figure \ref{gamma c}.

\par On the other hand, neutral Hayward black holes fails to demonstrate the onset of chaos under temporal perturbations simply because the required nonlinearity is suppressed by the lack of charge-dependent variables in the equation of state. This finding is in agreement with previous assessment and illustrates how charge serves a key part in driving thermodynamic instability towards chaotic behavior. This leads to the general condition that the equation of state of the black hole must have a higher power of the specific volume than a critical threshold in order for chaos to occur. This requirement is naturally met in the case of charged Hayward black holes with string fluids because of the combined influence of the nonlinear matter content and charge $q$.

\par Further, extending the methodology applied in the analysis of temporal chaos and using the ambient pressure parameter $B$ as a control variable in the perturbed thermodynamic system, we established a small spatially periodic thermal perturbation and obtained the corresponding dynamical equation. Upon investigating the qualitative behavior of the system for different values of ambient pressure $B$ relative to the phase transition pressure $P_0$, we constructed representative phase portraits corresponding to three distinct dynamical regimes, as shown in Figures~\ref{case 1}, \ref{case 2} and \ref{Case 3}. In each phase representation, the structure of homoclinic and heteroclinic orbits in the phase space is defined by a central center-type equilibrium point $v_2$, which is enclosed by two saddle points $v_1$ and $v_3$. These trajectories define the limits of stability in the spinodal region and encapsulate the system's nonlinear response to spatial perturbations. Such topological structures in the dynamical flow are a strong sign of chaotic dynamics and qualitatively align with observations in other AdS black hole backgrounds as well as van der Waals fluid systems that show phase transitions~\cite{Zhou:2022eft,Kumar:2024qon,MahishBhamidipati2019,SLEMROD1985135,Chabab:2018lzf,Barzi:2024bbj}.

A complementary perspective on the dynamics is provided by the Lyapunov exponents of circular null and timelike geodesics. Unlike the Melnikov method, which detects the global onset of chaos in the perturbed thermodynamic phase space through the splitting of homoclinic or heteroclinic structures, the Lyapunov exponent measures the local instability of geodesic motion. It therefore allows us to track how the dynamical instability changes across different thermodynamic branches and with the black-hole parameters. In particular, for fixed charge, we find that increasing the string fluid parameter $a$ lowers the Lyapunov exponent in both the null and timelike sectors. This shows that the string fluid suppresses local orbital instability and hence has a stabilizing effect on the black hole dynamics. This behavior is consistent with the Melnikov analysis, where $a$ also enters the critical perturbation threshold and controls the sensitivity of the system to the onset of chaos. Taken together, the Melnikov and Lyapunov analyses show that the string fluid parameter influences both the global chaotic structure of the perturbed thermodynamic phase space and the local instability of the underlying geodesic motion.


\quad In a nutshell, we have investigated the onset of chaos under temporal and spatial periodic perturbations in the extended thermodynamic phase space of Hayward-AdS black holes with string fluids. Using Melnikov's approach, we have shown that the nonlinear dynamics responsible for chaotic behavior in the spinodal region is governed by the interplay between the charge $q$ and the string fluid parameter $a$. While spatial perturbations can generate chaotic trajectories even in neutral configurations, charged black holes exhibit clear signatures of chaos under temporal perturbations. Complementing this global phase-space analysis, our Lyapunov exponent study shows that the local instability of circular null and timelike geodesics also encodes the thermodynamic phase structure of the black hole. In particular, the behavior of the Lyapunov exponent across different branches and its dependence on the parameters $q$ and $a$ demonstrate that the string fluid not only affects the onset of chaos in the Melnikov analysis but also suppresses local geodesic instability. The results presented here therefore highlight how matter content and regularization parameters shape both the phase-space structure and the stability properties of regular black holes. Chaos and Lyapunov instability can thus be used as complementary diagnostic tools for probing the underlying structure of regular geometries and the effective matter fields supporting them. Beyond offering a phenomenological window into possible quantum-gravity effects, the presence or absence of chaos, together with its sensitivity to coupling parameters, may also provide useful insight into the microscopic structure underlying black hole thermodynamics.

\quad There are several intriguing avenues for future research that merit consideration: First, using techniques like OTOCs and butterfly velocity, we may study holographic duals of thermodynamic chaos in regular AdS black holes, which could enrich our knowledge of chaotic dynamics from the perspective of boundary field theory. Incorporating quantum corrections via quantum-corrected string fluids or loop quantum gravity, on the other hand, may provide insight into how quantum influences impact the onset of chaos. Third, dimensional dependencies in chaotic behavior and critical phenomena may be revealed by higher-dimensional generalizations of Hayward-type black holes with generalized matter fields. Finally, using quench techniques to study non-equilibrium dynamics may help connect the more general ideas of non-equilibrium statistical mechanics with black hole thermodynamics.

\section*{Acknowledgments}
A.S. is grateful to Sandip Mahish and Ankit Anand for insightful discussion. A.M. extends gratitude for the hospitality received while presenting this work at the Indian Strings Meeting 2025, at IIT Bhubaneswar. The work of A.S. is supported by CSIR-HRDG under Project No. 03WS(003)/2023-24/EMR-II/ASPIRE.  B.P. thanks CSIR-HRDG, Govt of India, for financial support received through Grant No. 03WS(003)/2023-24/EMR-II/ASPIRE.


\bibliographystyle{JHEP}
\bibliography{ref}
\end{document}